\documentclass[aps,prd,twocolumn,showpacs,superscriptaddress,nofootinbib,floatfix]{revtex4-1}  % for review and submission
\usepackage{graphicx}  % needed for figures
\usepackage{dcolumn}   % needed for some tables
\usepackage{bm}        % for math
\usepackage{comment}
\usepackage{amssymb}   % for math
\usepackage{amsmath}
\usepackage{tabularx}
\usepackage{hyperref}
\usepackage{tabulary}

\DeclareMathOperator\erf{erf}

\begin{document}

\preprint{APS/123-QED}

  \title{Binary Black Hole Mergers from Globular Clusters: Masses, Merger Rates, and the Impact of Stellar Evolution}
  \author{Carl L.\ Rodriguez}  
  \affiliation{Center for Interdisciplinary Exploration and Research in
    Astrophysics (CIERA) and Dept.~of Physics and Astronomy, Northwestern
      University, 2145 Sheridan Rd, Evanston, IL 60208, USA}
  \author{Sourav Chatterjee}
    \affiliation{Center for Interdisciplinary Exploration and Research in
    Astrophysics (CIERA) and Dept.~of Physics and Astronomy, Northwestern
      University, 2145 Sheridan Rd, Evanston, IL 60208, USA}

  \author{Frederic A.\ Rasio}
    \affiliation{Center for Interdisciplinary Exploration and Research in
    Astrophysics (CIERA) and Dept.~of Physics and Astronomy, Northwestern
      University, 2145 Sheridan Rd, Evanston, IL 60208, USA}

\date{\today}% It is always \today, today,

\begin{abstract}
The recent discovery of GW150914, the binary black hole merger detected by Advanced LIGO, has the potential to revolutionize observational astrophysics.  But to fully utilize this new window into the universe, we must compare these new observations to detailed models of binary black hole formation throughout cosmic time.  Expanding upon our previous work \cite{Rodriguez2015a}, we study merging binary black holes formed in globular clusters using our Monte Carlo approach to stellar dynamics. We have created a new set of 52 cluster models with different masses, metallicities, and radii to fully characterize the binary black hole merger rate.  These models include all the relevant dynamical processes (such as two-body relaxation, strong encounters, and three-body binary formation) and agree well with detailed direct $N$-body simulations.  In addition, we have enhanced our stellar evolution algorithms with updated metallicity-dependent stellar wind and supernova prescriptions, allowing us to compare our results directly to the most recent population synthesis predictions for merger rates from isolated binary evolution.  We explore the relationship between a cluster's global properties and the population of binary black holes that it produces. In particular, we derive a numerically calibrated relationship between the merger times of ejected black hole binaries and a cluster's mass and radius. With our improved treatment of stellar evolution, we find that globular clusters can produce a significant population of massive black hole binaries that merge in the local universe.  We explore the masses and mass ratios of these binaries as a function of redshift, and find a merger rate of $\sim 5$ Gpc$^{-3}$yr$^{-1}$ in the local universe, with 80\% of sources having total masses from $32M_{\odot}$ to $64M_{\odot}$.  Under standard assumptions, approximately 1 out of every 7 binary black hole mergers in the local universe will have originated in a globular cluster, but we also explore the sensitivity of this result to different assumptions for binary stellar evolution. If black holes were born with significant natal kicks, comparable to those of neutron stars, then the merger rate of binary black holes from globular clusters would be comparable to that from the field, with approximately $1/2$ of mergers originating in clusters. Finally we point out that population synthesis results for the field may also be modified by dynamical interactions of binaries taking place in dense star clusters which, unlike globular clusters, dissolved before the present day.
\end{abstract}

\maketitle

%\tableofcontents

\section{Introduction}
\label{sec:intro}

With the recent detection of the binary black hole merger GW150914 \cite{Abbott2016a}, the era of gravitational-wave (GW) astronomy is here.  Advanced LIGO promises to open an
entirely new window into the universe, performing detailed measurements of
cataclysmic events such as the mergers of compact objects.  In
particular, the mergers of binary neutron stars (NS) and binary
black holes (BHs) will help answer detailed questions about such diverse fields
as stellar evolution, gamma-ray bursts, the NS
equation-of-state, and even general relativity itself \cite{Read2009,Cornish2011,Nissanke2010,Stevenson2015}.  But to fully realize the science potential of these
instruments, we must be prepared to relate GW observations
to detailed models of the compact object merger rate throughout the
universe.

To that end, significant work has been done to translate our understanding
of star formation and stellar evolution into realistic predictions for compact
object merger rates.  Population synthesis codes \cite[e.g.][]{PortegiesZwart2001a,Spera2015a,Belczynski2002,Belczynski2008}
use simple recipes for stellar evolution to rapidly model the evolution of a large population of stellar binaries.  However, this approach suffers from significant uncertainties.  Only weak observational constraints for binary NS systems exist from observations of binary pulsars in the Milky Way
\cite{Kalogera2004}, while the best rate estimates for stellar-mass binary BH (BBH) mergers arise from only one GW detection (GW150914) and one less-significant GW trigger (LVT151012) \cite{Abbott2016b}.  The prescriptions used to
predict such systems from population synthesis methods are tuned to observations of stellar wind-mass loss rates and low-mass X-ray binaries \cite{Belczynski2010a,Dominik2012}
which are few and far between, particularly for the low-metallicity massive stars that will
potentially dominate the BBH merger rate \cite{Belczynski2010}.  Furthermore, uncertainties in the physics of binary stellar evolution make it difficult to constrain the BBH merger rate from isolated binaries to within several orders of magnitude.

%Perhaps the most significant uncertainty, however, is the effect of dynamics on the potential population of BBHs.  When population synthesis codes model the star formation history of the universe, they typically neglect the observational result that \emph{most star formation occurs in clusters} \cite{Lada2003}.  While most of these clusters will disperse by the present day (becoming what we call the field), even the short lifetimes of young star clusters ($\lesssim 10^8$ yr) allow for BBHs to undergo multiple dynamical encounters before merger \cite[e.g.][]{Ziosi2014}.

In \cite{Rodriguez2015a}, we explored the contribution to the
BBH merger rate from globular clusters (GCs), a population of dense, old stellar
systems observed in most galaxies.  BBH formation in GCs is primarily driven by dynamics, with three-body binary
formation and strong gravitational encounters creating the
majority of BBH systems.  Since these clusters have survived to the present day,
they do not suffer from the uncertainties in star formation history, and present
a population of systems with well-measured masses and metallicities.
Furthermore, because most BBHs in GCs are formed dynamically, they are less sensitive to the physics of binary stellar evolution.   We
analyzed a collection of GC models from \cite{Morscher2015}, and found that a
single Advanced LIGO observatory could detect $\sim$10 to $\sim$100 BBH mergers from GCs per year, similar to many estimates of the merger rate from the field; however, these models assumed a distribution of BH masses that made it difficult to directly compare our results to population synthesis studies.

In this paper, we explore the full range of masses, mass ratios, eccentricities, and merger rates of BBH systems created in GCs.  To compare our results to recent estimates from the field, we have upgraded the stellar evolution algorithms of our dynamical models with new prescriptions for the metallicity-dependent stellar winds of O/B stars \cite{Vink2001} and a new prescription for the remnant and fallback masses of compact objects after supernova \cite{Fryer2012}.  These changes bring our code into agreement with the \texttt{StarTrack} population synthesis code \cite{Belczynski2002,Belczynski2008}, and allow us to create GC models with the same BH mass distribution used recent estimates for the merger rate from the galactic field \cite{Dominik2012,Dominik2013,Dominik2014}.  With these new stellar-wind prescriptions, we now find that GCs can produce significant numbers of massive BBHs that merge in the local universe, including $30M_{\odot}+30M_{\odot}$ BBHs, similar to GW150914.

In Section \ref{sec:cmc}, we describe the basics of our Monte Carlo approach to
modeling dense star clusters, our choice of initial conditions, and the upgrades
to our stellar evolution algorithms.  In Section \ref{sec:binprop}, we explore
how the global properties of a GC determine the masses, inspiral times, and
merger rates of the BBH population.  Then in Section \ref{sec:bbhprop}, we show
the distributions of merging BBH masses, mass ratios, and eccentricities as a
function of redshift.  We also explore the effects of various assumptions in our
stellar evolution algorithm, such as differences in the common-envelope physics,
the supernova mechanism, and the natal kicks given to BHs at formation.
Finally, in Section \ref{sec:rates}, we calculate the merger rates for BBHs from
GCs as a function of redshift, and compare these rates to the most recent
estimates of merger rates from the field \cite{Dominik2013}.  Throughout this
paper, we assume cosmological parameters of $\Omega_M = 0.3$, $\Omega_\Lambda =
0.7$, and $H_0 = 70$km s$^{-1}\,\text{Mpc}^{-1}$, to compare with the most recent estimates of BBH mergers from galactic fields \cite{Dominik2013}.

\section{Cluster Models}
\label{sec:cmc}

We create 52 GC models with our Cluster Monte Carlo code (CMC), an orbit-sampling H\'enon-style Monte Carlo code \cite{Henon1971a,Henon1971}.  CMC has been developed and described in several previous papers \cite{Joshi1999,Joshi2001a,Fregeau2003,Fregeau2007,Chatterjee2010,Pattabiraman2013}.  By assuming that the dynamics of a GC is primarily driven by two-body relaxation between neighboring particles, CMC can create models of dense star clusters on a significantly faster timescale than a traditional direct $N$-body integrator.  This allows us to fully explore the parameter space of GCs, including the massive and compact clusters used in this study.  Recent comparisons to state-of-the-art $N$-body simulations \cite{Wang2015} have shown that CMC can correctly model the evolution of realistic clusters with $\sim10^6$ particles over 12 Gyr, including the masses, semi-major axes, and eccentricities of the ejected BBHs \cite{Rodriguez2016}.

In addition to two-body relaxation, CMC incorporates all the relevant physics for treating the BBH formation and merger problem.  This includes:

\begin{itemize} 

\item binary-single and binary-binary gravitational encounters computed
explicitly with the \texttt{Fewbody} small-$N$ integrator \cite{Fregeau2004},

\item three-body binary formation, implemented with an analytic prescription \cite{Morscher2012}, and

\item single and binary stellar evolution, implemented with BSE \cite{Hurley2000,Hurley2002}, and improved with the stellar remnant and BH kick prescription from \cite{Kiel2009,Chatterjee2010}.  For this paper, we also update the prescriptions for stellar winds and supernova fallback, in order to replicate the BH mass distribution of \cite{Belczynski2010,Dominik2012,Dominik2013,Dominik2014} (see Section \ref{subsec:newse})

\end{itemize}

The BBH merger time is found by directly integrating the orbit-averaged Peters equations \cite{Peters1964}.  For mergers in the cluster, this is calculated by BSE.  For mergers of BBHs that were ejected, we calculate the inspiral time from the masses, semi-major axis, and eccentricity at ejection.  The merger time is then the sum of the inspiral time and the time at which the binary is ejected from the cluster.

\begin{table*}[t]
\begin{tabular*}{\textwidth}{@{\extracolsep{\fill}} ccc|ccc}
\hline\hline
\multicolumn{3}{c}{\textbf{Initial Conditions}} & \multicolumn{3}{c}{\textbf{Final Properties (12 Gyr)}} \\\hline
N ($\times 10^5$) & $R_v$ (pc) & Metallicity $(Z_{\odot})$& Mass ($10^5 M_{\odot}$)& $N_{\text{bin}}$ & $N_{\text{insp}}$ \\ \hline

2 & 1 & 0.01 & 0.51, 0.52 & 36, 44 & 9, 7\\
2 & 1 & 0.05 & 0.46, 0.43 & 41, 44 & 5, 10\\
2 & 1 & 0.25 & Dissolved & -- & --\\\hline

5 & 1 & 0.01 & 1.47, 1.48 & 90, 89 & 29, 32\\
5 & 1 & 0.05 & 1.37, 1.36 & 87, 86 & 28, 33\\
5 & 1 & 0.25 & 0.85, 0.85 & 80, 84 & 22, 25\\\hline

10 & 1 & 0.01 & 3.05, 3.07 & 177, 169 & 90, 75\\
10* & 1 & 0.05 & 2.84, 2.94 & 178, 158 & 94, 84\\
10 & 1 & 0.25 & 2.35, 2.35 & 178, 168 & 102, 86\\\hline

20 & 1 & 0.01 & 6.32, 6.32 & 321, 334 & 232, 233\\
20 & 1 & 0.05 & 5.96, 5.95 & 339, 334 & 254, 252\\
20 & 1 & 0.25 & 5.29, 5.33 & 323, 348 & 260, 270\\\hline\hline

2 & 2 & 0.01 & 0.4, 0.5 & 31, 32 & 3, 4\\
2 & 2 & 0.05 & 0.47, 0.48 & 37, 31 & 7, 6\\
2 & 2 & 0.25 & Dissolved & -- & --\\\hline

5 & 2 & 0.01 & 1.49, 1.52 & 82, 75 & 22, 21\\
5 & 2 & 0.05 & 1.4, 1.41 & 74, 75 & 21, 23\\
5 & 2 & 0.25 & 0.72, 0.73 & 73, 71 & 20, 15\\\hline

10 & 2 & 0.01 & 3.18, 3.15 & 121, 127 & 43, 48\\
10 & 2 & 0.05 & 3.0, 3.01 & 133, 140 & 55, 54\\
10 & 2 & 0.25 & 2.53, 2.47 & 128, 136 & 44, 46\\\hline

20 & 2 & 0.01 & 6.59, 6.59 & 215, 233 & 144, 139\\
20 & 2 & 0.05 & 6.33, 6.3 & 243, 226 & 154, 153\\
20 & 2 & 0.25 & 5.86, 5.82 & 221, 230 & 132, 127\\\hline\hline

\end{tabular*}
\caption{List of the main 48 GC models created in this study.  We consider 24 initial conditions with different initial particle numbers, virial radii, and metallicity, and run two independent realizations of each cluster model.  For each set of model inputs, we show the final cluster mass, the total number of BBHs ejected by 12 Gyr ($N_{\text{bin}}$), and the number of those BBHs that will merge before the present day ($N_{\text{insp}}$) for each of the two models.  Of these 48 models, 4 dissolve before 12 Gyr.  We show these for completeness, but do not include them in our analysis, which we have restricted to the population of GCs that survive to the present day.  We also generate 4 additional models, identical to the starred model, with different physics for the treatment of stellar evolution.  These are described in Section \ref{subsec:se}.}
\label{tab:models}
\end{table*}

\subsection{Initial Conditions}
\label{subsec:ic}

For our main result, we create 52 GC models spanning a range of
different realistic parameters.  These are described in Table \ref{tab:models}.  We consider a 3x2x4 grid of models, with different metallicities, virial radii, and initial particle numbers, for a total of 24 initial conditions, and four additional models in which we varying the assumptions of stellar evolution.  For each of the 24 main models, we generate two statistically independent cluster models for each initial condition.  We
consider three metallicities: $0.01 Z{\odot}$, corresponding roughly to
the lower-end of the Milky Way GC (MWGC) metallicity distribution \cite{Harris1996}, $0.05 Z_{\odot}$,
corresponding to the peak of the MWGC distribution, and $0.25
Z_{\odot}$, corresponding to the peak of the high-metallicity MWGCs.  We will refer to our $0.01Z_{\odot}$ and $0.05Z_{\odot}$ models as low-metallicity, and our $0.25Z_{\odot}$ model as high-metallicity\footnote{ GCs in most galaxies are observed to fall into two distinct metallicity groups: high-metallicity, or ``red'' GCs, which are typically found in the galactic bulge, and low-metallicity, or ``blue'' GCs, typically found in the galactic halo \cite{Harris2010}.}.  We place
the $0.25Z_{\odot}$, $0.05Z_{\odot}$, and $0.01Z_{\odot}$ clusters at galactocentric distances of 2 kpc,
8 kpc, and 20 kpc respectively, as there exists a strong correlation between a
cluster's galactocentric distance and its metallicity \cite{Djorgovski1994}.
We also consider clusters with initial virial radii of both $1\,\rm{pc}$ and
$2\,\rm{pc}$.
This is motivated by both observational evidence of cluster formation in the
local universe, suggesting that $R_v \sim 2\,\rm{pc}$ is typical \cite{Scheepmaker2007} and theoretical modeling suggesting
that smaller $R_v$ produce GCs with properties more similar to observed MWGCs
\cite{Morscher2015}.  

For each metallicity, galactocentric distance, and virial radius, we then
consider clusters with $N=2\times10^5$, $5\times10^5$, $1\times10^6$, and
$2\times10^6$ particles.  The particles are placed at radii drawn
from a King profile \cite{King1966} with $W_0 = 5$ (as
\citep{Morscher2015,Rodriguez2015a} found no correlation between the initial GC
concentration and either the observational properties of the clusters or the properties of the BBH population).  The stellar masses are drawn from a Kroupa
initial mass function (IMF) of the form \cite{Kroupa2001a}:

\begin{equation}
P(m)dm \propto m^{-\alpha}dm
\label{eqn:kroupa}
\end{equation}

\noindent with

\begin{equation}
  \alpha=\begin{cases}
  1.3 & 0.08M_{\odot} \leq M < 0.5 M_{\odot} \\ 
  2.3 & 0.5 M_{\odot} \leq M
  \end{cases}
\end{equation}

\noindent We consider masses from $0.08M_{\odot}$ to $150M_{\odot}$, in order to
directly compare our results to recent results for the field (e.g.~\cite{Dominik2012,Dominik2013}). We also assume that all models begin with 10\%
of the particles as binaries.  This is accomplished by randomly selecting 10\% of
particles and adding a companion mass drawn from a flat distribution in the mass ratio.  The binary separations are drawn from a $P(a)da \propto 1/a$ distribution, with a lower-limit near the point of stellar contact and an upper-limit such that the velocity of the binary components is equal to the average velocity of a star in the cluster core.
The eccentricities are chosen from a thermal
distribution, $P(e)de = 2e de$ \cite{Heggie1975}.  Because we select 10\% of our
particles to become binaries, our models have 10\% more stars than the number of
particles (e.g. our $N=2\times10^5$ model contains $2.2\times10^5$ stars).

\subsection{Upgraded Stellar Evolution}
\label{subsec:newse}

In \cite{Rodriguez2015a}, we used a series of models from \cite{Morscher2015} to estimate the BBH merger rate.  We noted that in our simulations, dynamically-formed BBHs from GCs were characteristically more massive than those formed by stellar evolution of binaries in the field; however, this result was highly dependent on our particular prescription for binary stellar evolution.  More recent work \cite{Belczynski2010,Dominik2012,Spera2015a} has suggested that massive BBHs (with total masses up to 160 $M_{\odot}$) can be formed in low-metallicity environments, where the reduced stellar winds from massive stars enable the formation of BHs with masses up to $80 M_{\odot}$.  Furthermore, the reduced wind-driven mass loss increases the chance that the system will remain bound during its evolution to become a BBH.  

We have upgraded the binary stellar evolution module in CMC with
improved prescriptions for the metallicity-dependent stellar winds and
supernova-driven mass loss.  For stellar winds, we make two additions to the
default BSE implementation: first, for O and B stars, we implement the
mass-loss prescriptions developed in \cite{Vink2001}, which we will refer to as
the ``Vink prescription''.  This fit determines the mass-loss as a function of the
star's mass, effective temperature, metallicity, and luminosity for any hot
($12500\,\rm{K} < T_{\text{eff}} < 50000\,\rm{K}$) hydrogen-rich star on the main sequence.
The Vink prescription covers separately the temperature range from $12500\,\rm{K} <
T_{\text{eff}} < 22500\,\rm{K}$ and $27500\,\rm{K} < T_{\text{eff}} < 50000\,\rm{K}$, while
explicitly excluding the range between $22500\,\rm{K}$ and $27500\,\rm{K}$, which is
complicated by the appearance of Fe ion line-driven winds at approximately
$25000\,\rm{K}$.  We follow the prescription from \cite{Belczynski2010a}, extending the
lower temperature prescription (equation (25) from \cite{Vink2001}) up to
$25000\,\rm{K}$ and the higher temperature prescription (equation (24) from \cite{Vink2001}) down to $25000\,\rm{K}$.  In addition to the new prescription for O and B stars, we add a metallicity dependence to the evolution of naked-helium (Wolf-Rayet) stars.  This is done by supplementing the original BSE prescription \cite{Hamann1998} with the metallicity dependence from \cite{Vink2005}.

We also include two new prescriptions for the supernova mechanism.  These prescriptions, first
developed in \cite{Fryer2012}, describe the amount of material that falls back onto the newly-formed compact object for
neutrino-driven and convection-enhanced supernovae.  They consider two cases,
based on the delay between the core bounce and the explosion: the
\textit{rapid} case, which assumes that any explosion occurs within the first 0.25
seconds after the core bounce, and the \textit{delayed} case, which relaxes this
assumption.  The rapid prescription replicates the observed ``mass-gap'' between
NSs and BHs \cite{Farr2011,Belczynski2012a}, while the delayed prescription
allows for BH masses from $2M_{\odot}$ to $5M_{\odot}$.  For the default in this
study, we use the rapid supernova model, although in Section \ref{subsec:se} we also explore the effects of the delayed supernova mechanism on BBH production in GCs

In addition to determining the mass of the supernova remnant, the mass of the fallback material determines the velocity kick the newly-formed BH receives from the explosion.  Consistent with the original version of BSE, we give
all BHs natal kicks drawn from a Maxwellian with a dispersion of $\sigma= 265$km s$^{-1}$ \cite{Hobbs2005} as is done for NSs formed by core-collapse supernova.  However, we assume that the final velocity is lowered by the mass of
the fallback material according to

\begin{equation}
V_{\text{final}} = (1 - f_{\text{fb}}) V_{\text{natal}}
\label{eqn:kick}
\end{equation}

\noindent where $f_{\text{fb}}$ is the fraction of the ejected supernova mass that will fall back onto the newly-formed proto-compact object, determined by equations (16) and (19) in \cite{Fryer2012}.  For any sufficiently massive BH progenitor ($M\gtrsim 40 M_{\odot}$), the fallback completely damps any natal kick, and the BH is retained in the cluster.  This is consistent with the direct collapse scenario discussed in \cite{Fryer2001}.

The goal of these modifications is to replicate the BH mass spectrum used in the
recent studies of field-born BBHs, specifically those employed in the \texttt{StarTrack} population synthesis code.  With these new changes to our version of
BSE, we find that CMC produces a similar relationship between the zero-age
main-sequence mass of a massive star and its final BH mass 
(Figure 11 of \cite{Fryer2012}).  Therefore we can now explicitly compare our
results to those from \cite{Dominik2013,Dominik2014}.  Note that this does not
imply that our stellar evolution code would replicate the \texttt{StarTrack}
results for binary stars, as we have not modified our binary stellar evolution prescription
from the version used in previous papers.  However, since the majority of BBHs
from GCs are formed dynamically (90\% of all BBH mergers, and 99.7\% of mergers at $z<1$), this does not
significantly impact our results.  To confirm this, we create an additional 4 GC
models varying the physics of binary stellar evolution, and explore the effects of these assumptions
in Section \ref{subsec:field}.  

\section{BBH Production in GCs}
\label{sec:binprop}

\begin{figure*}[tbh]
\centering
\includegraphics[trim=3cm 0cm 3cm 0cm,scale=0.6]{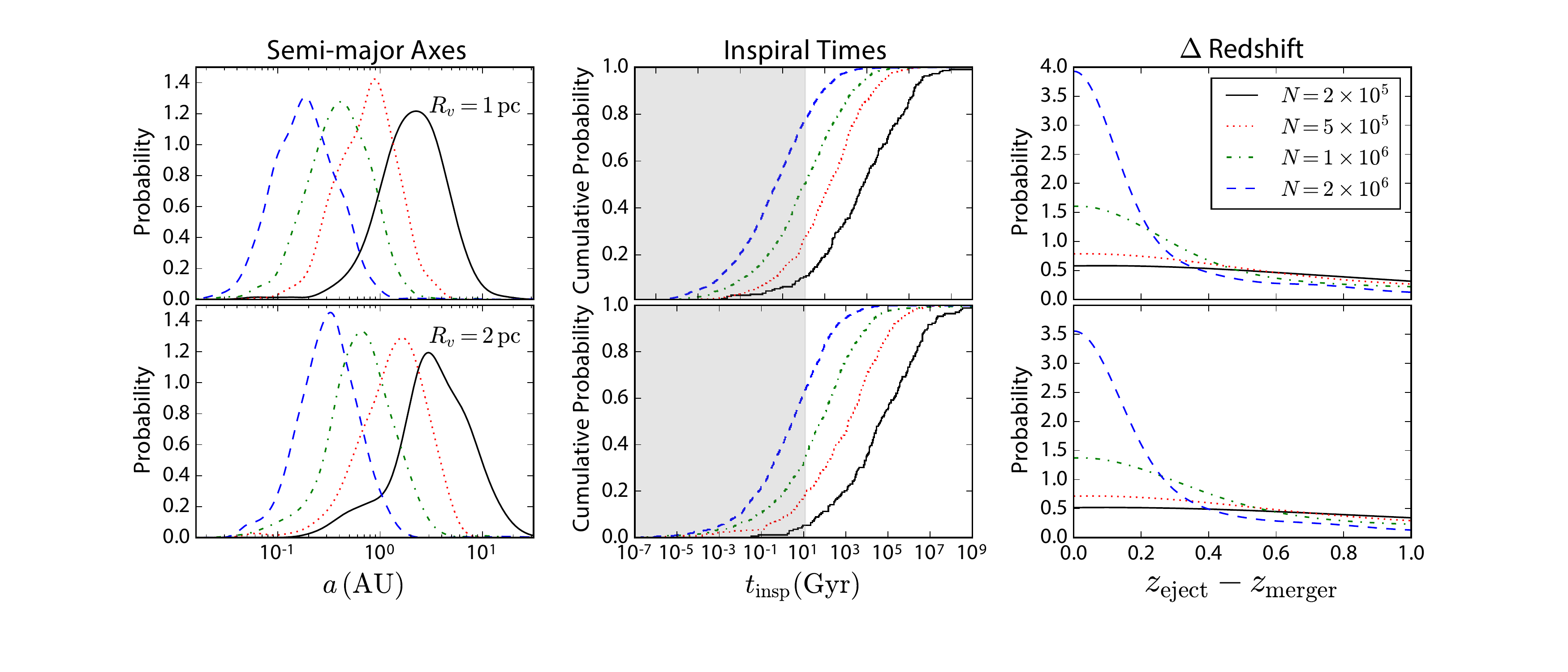}
\caption{The distribution of semi-major axes and inspiral times for the binaries
ejected from our collection of GC model.  The models are broken down according
to the initial number of particles (equivalent to the mass), with the $R_v =
1\,\rm{pc}$ models along the top row, and the $R_v= 2\,\rm{pc}$ models along the bottom.
Clusters with larger masses and smaller virial radii eject binaries with higher binding energies.  The gray shaded region in the plot of inspiral times indicates merger times of less that 12 Gyr.  The rightmost plots shows the differences between the redshift at which the binary is ejected from the cluster and the redshift at which it merges.}
\label{fig:inspirals}
\end{figure*}

\begin{figure}[tbh]
\centering
\includegraphics[trim=4cm 0cm 3cm 0cm,scale=0.78]{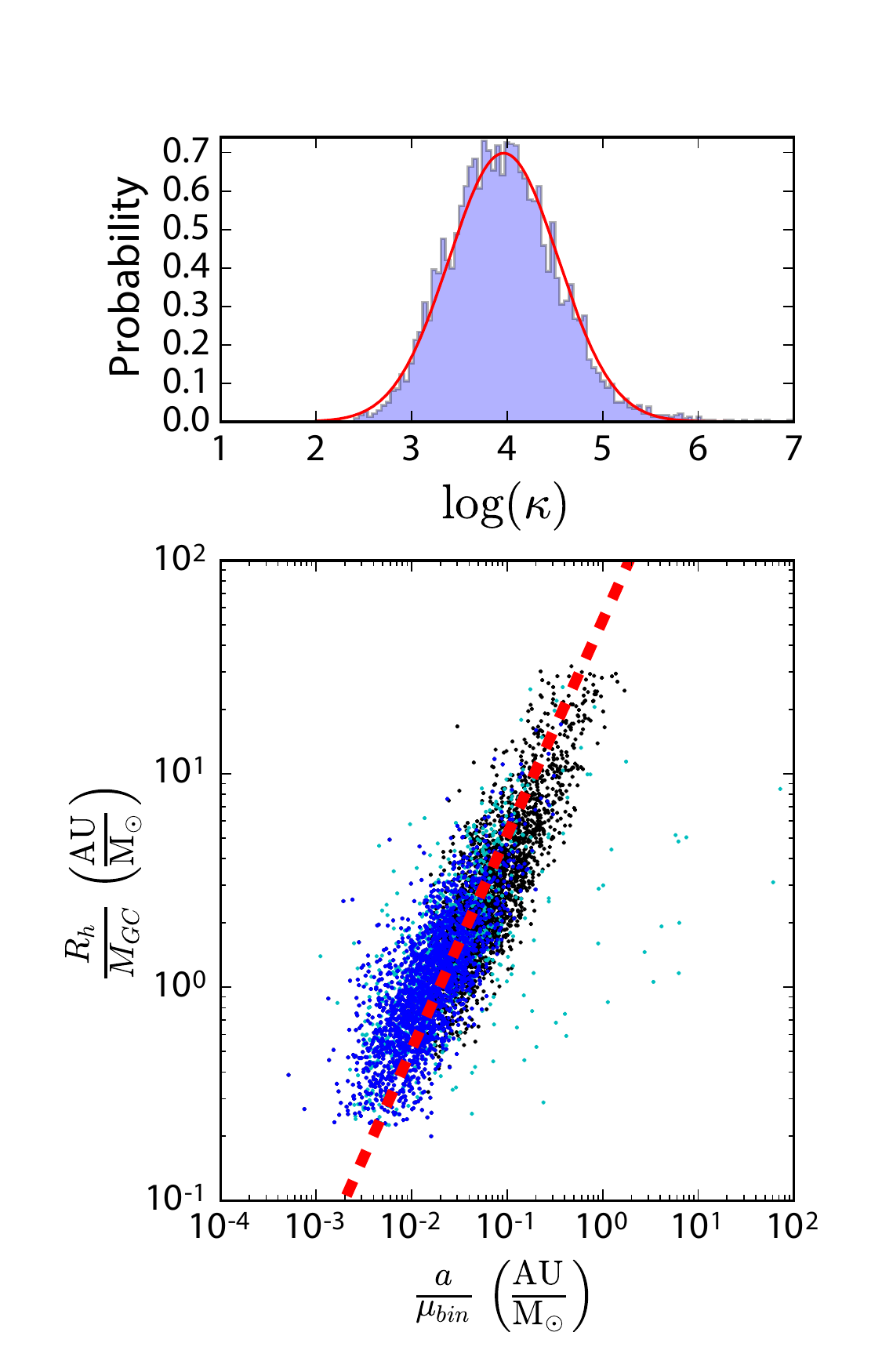}
\caption{The relationship between the global cluster parameters and the properties of its ejected binaries.  Clusters with higher escape speeds (low $R_h$ and high $M_{GC}$) eject binaries with larger binding energies.  The top panel shows the distribution of the constant from equation \eqref{eqn:semi-major}, fitted to a log-normal distribution.  The lower panel shows the relationship between the ratio of the semi-major axis to the reduced mass for each binary, $a/\mu_{\text{bin}}$, and the ratio of the half-mass radius to cluster mass, $R_h/M_{{GC}}$, at the time each binary is ejected from the cluster.  The red dashed line indicates the median of the log-normal fit.  The blue and black points show each binary ejected after a strong encounter with a single star, with the blue points indicating binaries that will merge within 12Gyr.  The cyan points (excluded from the log-normal fit) show binaries ejected after a strong encounter with another binary.}
\label{fig:semi-major}
\end{figure}

\begin{figure}[tbh]
\centering
\includegraphics[trim=2cm 0cm 2cm 0cm,scale=0.75]{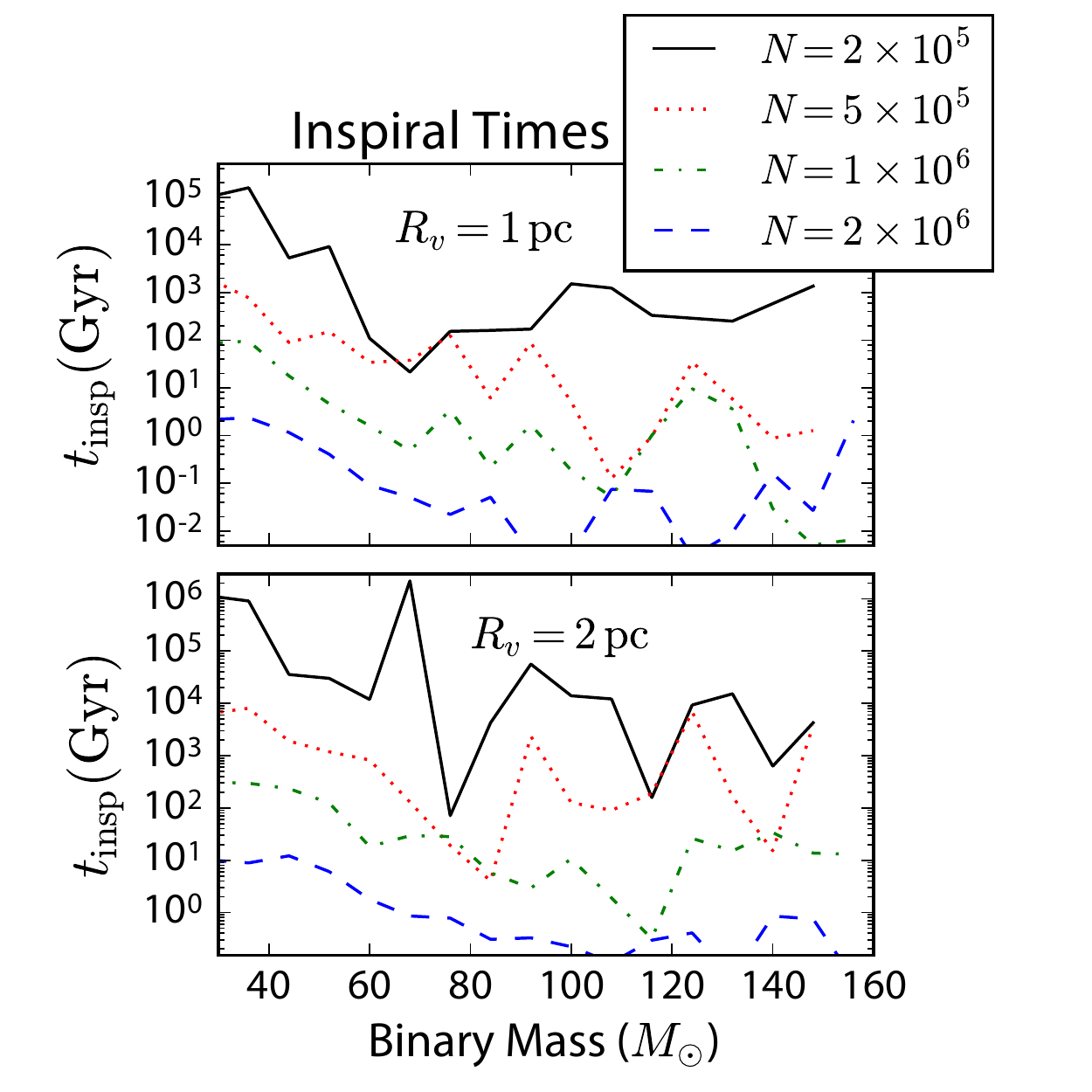}
\caption{The median inspiral time for binaries from different GCs as a function of binary mass.  The binaries from each cluster are binned according to total mass, and the median inspiral time (after ejection from the cluster) is computed.  An increase in either the binary mass or the cluster mass yields a decrease in the median merger times.  The fluctuations in each line arise from the small number of points, which more strongly effect high-mass BBHs and smaller GC models.}
\label{fig:insp_times}
\end{figure}

\begin{figure}[tbh]
\centering
\includegraphics[trim=2cm 0cm 2cm 0cm,scale=0.75]{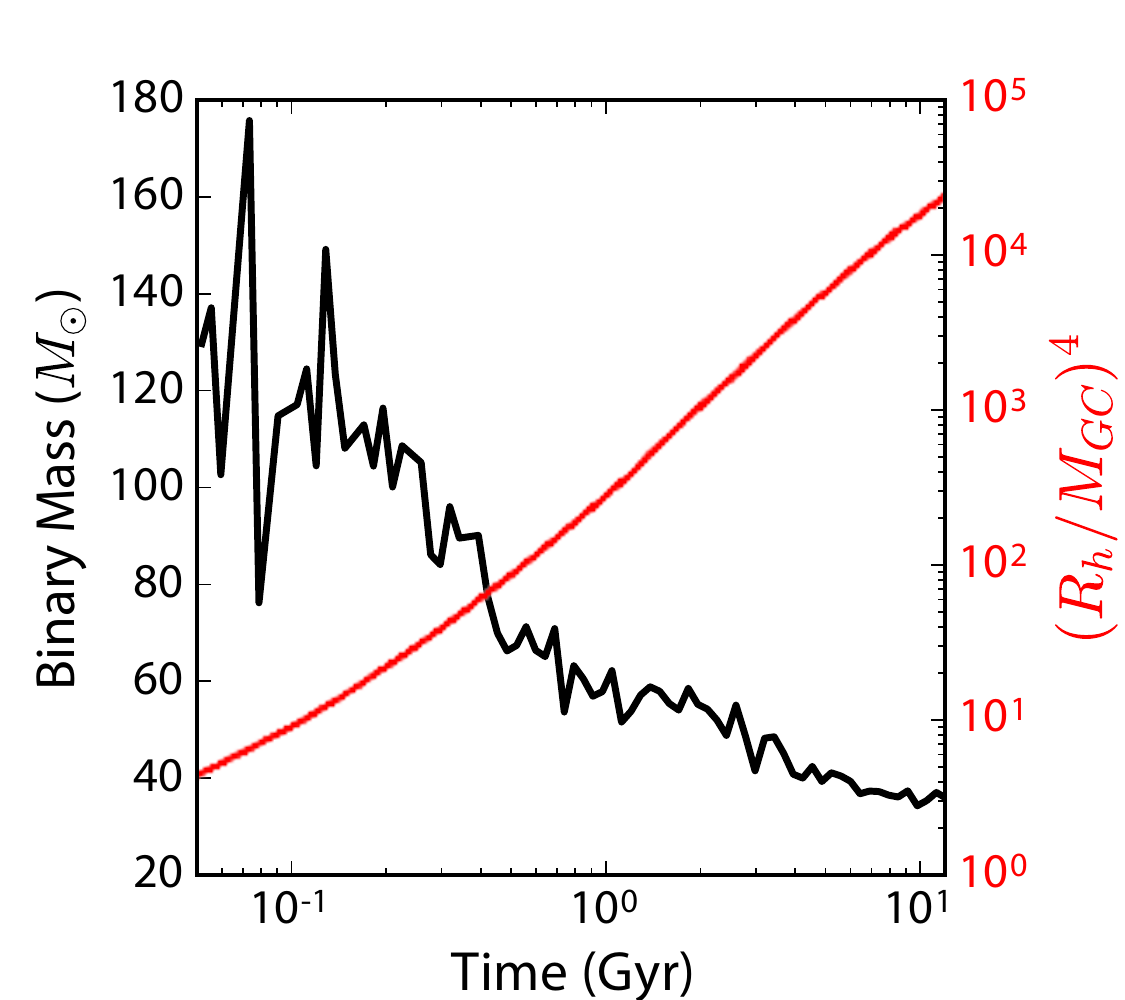}
\caption{The relationship between the median mass of an ejected BBH (in black)
and the ratio of the half-mass radius to the GC mass (in red) as a function of
time for our $0.01Z_{\odot}$, $N=2\times10^6$, $R_v=1\,\rm{pc}$ model.  The cluster preferentially forms and ejects massive binaries early in its lifetime,  However, during this phase, the $(R_h / M_{GC})^4$ term increases by several orders of magnitude, primarily driving the increase in inspiral times as shown in equation \eqref{eqn:tinsp_m}.  We normalize $R_h$ and $M_{GC}$ to their initial values at $t=0$.}
\label{fig:insp_vs_rcm}
\end{figure}

\subsection{Binary Properties}
\label{subsec:binprop}

One of the most interesting results of \cite{Rodriguez2015a} was the significant
increase in the merger rate of BBHs from GCs as compared to other studies.  This was
attributable to two key effects: first, the Monte Carlo method allows us to
model significantly larger clusters than a direct
$N$-body method.  Since CMC can integrate clusters with $N \sim 10^6$ in days to weeks, \cite{Rodriguez2015a} was able to explore the contribution of massive GCs that were shown to dominate the BBH merger rate.  Previous studies, particularly those using
direct $N$-body integrators, were limited to clusters with at most $\sim10^5$
particles \cite{Zwart1999,Banerjee2010,Tanikawa2013,Bae2014}.  After 12 Gyr of evolution, these clusters would
produce at most $\sim$10 to 20 BBHs, only $\sim 10\%$ of which would merge in a Hubble
time.  The one exception was the two studies
\cite{Downing2010,Downing2011}, which used a Monte Carlo approach similar to CMC.  However, they only considered a single, relatively low cluster mass ($N \sim
5\times10^5$) and  did not include direct
integration of binary-single and binary-binary encounters, critical for
correctly predicting the properties of dynamically-formed BBHs.  By covering the full distribution of GC masses, our models can correctly predict
the number of binaries ejected from the most massive GCs, significantly
increasing the total number of binaries produced by GCs overall. 

Not only do massive clusters produce more BBHs, but the BBHs that they produce
are \emph{more likely} to merge within 12 Gyr than BBHs from lower-mass GCs.  This result arises from the physics of binary
formation and interaction in a cluster environment.  A
hard binary (with binding energy greater than the typical kinetic
energy of particles in the cluster),  typically
undergoes a series of strong encounters with other single and binary stars.  The
result of these repeated encounters, known as ``Heggie's law'', is that hard
binaries tend to get harder, increasing their binding energy on average with each encounter, and transferring that energy
to the center-of-mass velocity of the interacting stars
\cite{Heggie1975}.  The recoil velocity of the binary in the cluster is increased after
each encounter, until eventually the binary receives a velocity kick sufficient
to eject it from the cluster. 

Since conservation of energy demands that the change in $E_{\text{bin}}$ be proportional to the post-encounter kinetic energy of the binary, and the potential of the cluster determines the minimum kinetic energy for escaping particles, we can use the cluster potential to determine the minimum binding energy (and maximum semi-major axis) required to eject a binary.  First, we assume that $E_{\text{bin}}$ increases by roughly the same fraction after every binary-single encounter, and that $\sim1/3$ of that energy gets transferred into the binary recoil.  \cite{Heggie1975,Hut1992} both assumed an average $E_{\text{bin}}$ increase of 40\% after each encounter, while \cite{Zwart1999} assumed 20\%; however, both studies considered the average only for particles with equal masses.  For generality, we simply assume that the binding energy is related to the center-of-mass kinetic energy by some factor $\alpha$, such that after a scattering encounter 

\begin{equation}
E_{\text{CM}} = \alpha E_{\text{bin}} = \alpha \frac{G m_1 m_2}{2 a}
\label{eqn:ecm}
\end{equation}

\noindent where $a$ is the binary semi-major axis and $m_1$ and $m_2$ are the component masses.  If we assume the cluster potential is represented by a Plummer distribution, then the energy required to eject a binary from the center of the cluster is given by \cite[e.g.,][]{Heggie2003}

\begin{equation}
E_{\text{esc}} = (m_1 + m_2) | \Phi_c | = \frac{G M_{GC}(m_1 + m_2)}{\sqrt{2^{2/3}-1} R_h}
\label{eqn:ejec}
\end{equation}

\noindent where $M_{GC}$ is the cluster mass and $R_h$ is the half-mass radius (the radius enclosing half the cluster mass). Combining \eqref{eqn:ecm} and \eqref{eqn:ejec}, we can relate the semi-major axis of ejected binaries to the mass and half-mass radius of the cluster at the time of ejection by 

\begin{align}
 \frac{M_{GC}}{R_h} &= \left(\alpha \frac{ \sqrt{2^{2/3}-1}}{2}\right)\frac{\mu_{bin}}{a}\nonumber\\
  &= \frac{1}{\kappa} \frac{\mu_{bin}}{a}
  \label{eqn:semi-major}
\end{align}

\noindent where $\mu_{bin} = (m_1 m_2)/(m_1+m_2)$ is the reduced mass of the binary and we have defined the coefficient $\kappa \equiv 2/(\alpha\sqrt{2^{2/3}-1})$.  

Equation \eqref{eqn:semi-major} reinforces an important point: when averaged
over many encounters, it is the global properties of the cluster, not the
micro-physics of the dynamics, that primarily determines the
semi-major axes of binaries formed 
dynamically.  We show the semi-major axes, inspiral times, and
redshift delays for the binaries from each of our models in Figure
\ref{fig:inspirals}.  Although this relationship between the cluster potential and the
semi-major axis of ejected binaries has beeen noted before
\cite{Zwart1999,Moody2009}, the size of the models considered here allows us to
extend this analysis to realistically-sized clusters with realistic BBH masses.

Since our GC models span a range of BH masses, cluster masses, and cluster
half-mass radii, we can compare \eqref{eqn:semi-major} directly to the BBHs
ejected from our models.  In Figure \ref{fig:semi-major}, we show the
relationship between the semi-major axis and reduced mass of each binary and the
half-mass radius and total mass of the cluster from which it was ejected.  We
show the distribution of $\kappa$ from all ejected binaries in the top panel.
We find that the distribution of $\kappa$ values roughly follow a log-normal
distribution.   In the bottom
panel, we plot the value of $a/\mu_{bin}$ for every binary against $R_h/M_{GC}$
at the time it was ejected.  We also show a line corresponding to the median of
the log-normal fit.  We note that this relationship only applies to binaries
ejected from the cluster during strong encounters with a single object.
Binaries ejected during encounters with another binary have significantly more
complicated interactions, including a much greater possibility of exchanging
components.  However, we find that 81\% the of binaries ejected from our GC
models are ejected immediately following a binary-single encounter (involving
either a hardening encounter or an exchange of componenets), while only 13\% are ejected following a binary-binary encounter.  For completeness, we also show these points in Figure \ref{fig:semi-major}, though we exclude them from the fit to equation \eqref{eqn:semi-major}.  With the fit to $\kappa$ from equation \eqref{eqn:semi-major}, we can express the probability for a BBH to be ejected with a given eccentricity and semi-major axis as:

\begin{flalign}
&P(e)\,de = 2e\,de\label{eqn:eccen}\\
&P(a | M_{GC}, R_h, \mu_{\text{bin}})\,da = \frac{1}{a \sigma \sqrt{2\pi}} \label{eqn:pa}\times \\ &~~~~~~~~~~~~~~~~~~~~~~~~~~~~~\exp\left[ -\frac{\left(\log \frac{\mu_{\text{bin}} R_h}{a M_{GC}} - a^*\right)^2} {2\sigma^2} \right]\,da \nonumber
\end{flalign}

\noindent where $a^{*}$ and $\sigma$ are the parameters of the
log-normal distribution with mean $a^{*} = 3.98$ and $\sigma = 0.59$.  We reiterate that equations \eqref{eqn:eccen} and \eqref{eqn:pa} only apply for the 81\% of sources which are ejected from the cluster following a binary-single encounter.

With a relationship between the cluster parameters and the semi-major axes of the binaries it ejects, we can show how cluster dynamics determines the inspiral times of BBHs once they are ejected from the cluster.  From the Peters equations \cite{Peters1964}, we see that the merger time of a binary scales as

\begin{equation}
t_{\text{insp}} \propto \frac{a^4}{m_1 m_2 (m_1 + m_2)}
\label{eqn:peter_time}
\end{equation}

\noindent If we combine equation \eqref{eqn:peter_time} with the scaling from equation \eqref{eqn:semi-major}, we can show that the inspiral time of a binary with total mass $M\equiv m_1+m_2$ scales as

\begin{equation}
t_{\text{insp}} \propto \left( \frac{R_h}{M_{GC}} \right )^4 M
\label{eqn:tinsp_m}
\end{equation}

\noindent where we have assumed all binaries to have equal-mass components.

In Figure \ref{fig:insp_times}, we show the inspiral times for BBHs ejected from each of our GC models as a function of binary total mass.  What is immediately striking is that the median inspiral time appears to \emph{decrease} with increasing binary mass, in contrast to the scaling derived in equation \eqref{eqn:tinsp_m}.  This is primarily due to the influence of the cluster itself: while the binary mass does determine the inspiral time of the binary post-ejection, it is the cluster mass and half-mass radius that determine the separation at ejection, and the $(R_h / M_{GC})^4$ factor in equation \eqref{eqn:tinsp_m} increases by several orders-of-magnitude over the 12 Gyr lifetime of a GC.  

In Figure \ref{fig:insp_vs_rcm} we show the median mass of ejected BBHs and the ratio $(R_h / M_{GC})^4$ for a single GC model over time.  As the cluster ages, it preferentially ejects its most massive BHs first, working its way from most to least massive BHs as time progresses.  As the BHs are ejected, the cluster expands and loses mass, increasing $R_h / M_{GC}$ and significantly increasing the inspiral time of the lower-mass binaries ejected at late times.    This scaling drives the counter-intuitive decrease in inspiral times for high-mass BBHs seen in Figure \ref{fig:insp_times}: these BBHs are ejected early in the cluster lifetime, when $(R_h / M_{GC})^4$ is low.  As the cluster ages, $(R_h / M_{GC})^4$ increases significantly, and the average inspiral time for an ejected BBH increases accordingly.

\subsection{BBH Mergers per Cluster}
\label{subsec:mergerPerClus}

\begin{figure}[tb]
\centering
\includegraphics[trim=3cm 0cm 3cm 0cm,scale=0.8]{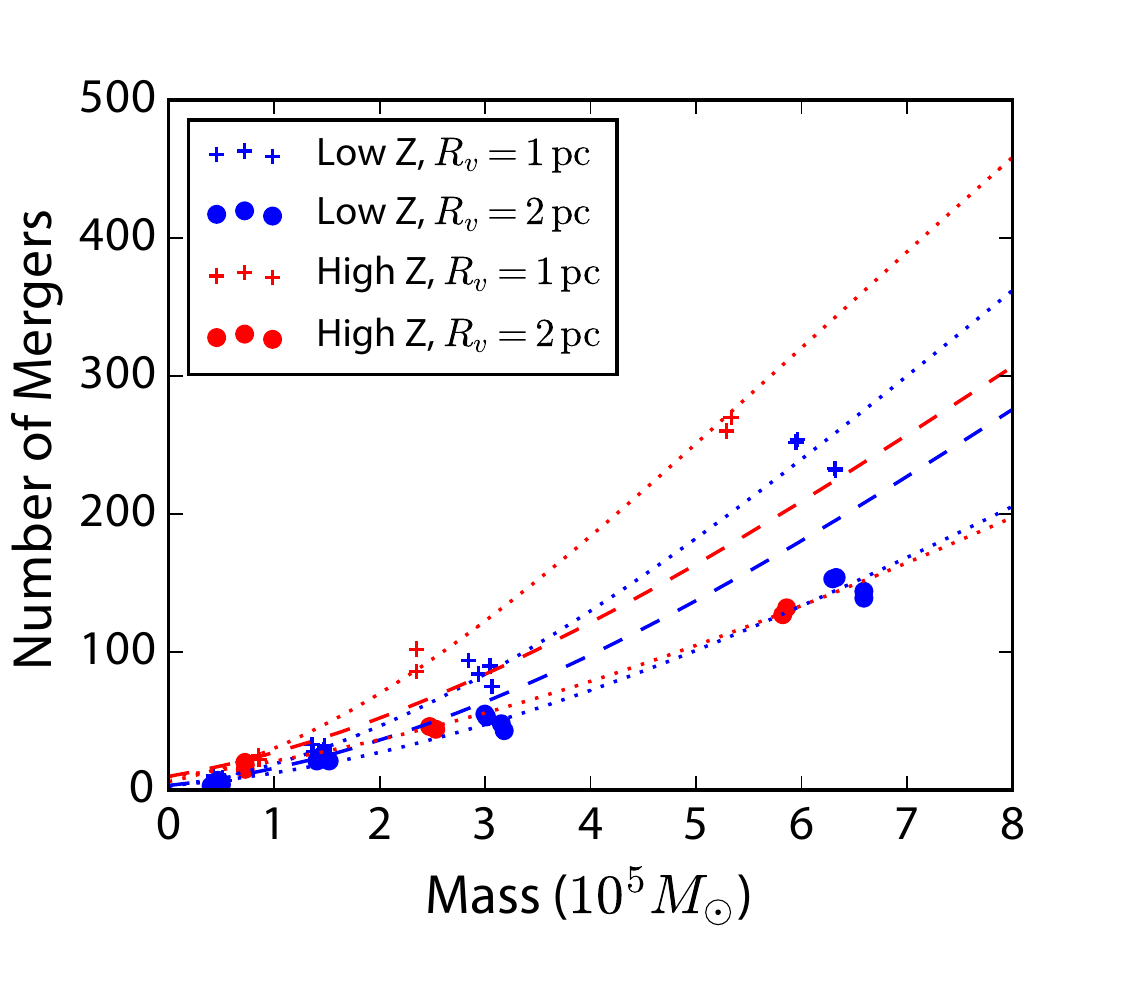}
\caption{The number of mergers for each model as a function of the final cluster
mass.  We show separately the low-metallicitiy ($0.01Z_{\odot}$,
$0.05Z_{\odot}$) and high-metallicity GCs ($0.25Z_{\odot}$), and separate the
clusters by virial radii.  We then fit equation \eqref{eqn:ninsp} to each set of
models.  The main fits using the combined sample of clusters with $R_v =
1\,\rm{pc}$ and $R_v = 2\,\rm{pc}$ are shown with dashed lines, while the dotted
lines show fits to clusters with only $R_v = 1\,\rm{pc}$ or $R_v=2\,\rm{pc}$.}
\label{fig:mergerPerMass}
\end{figure}

To determine the mean number of mergers per cluster, we need a functional form for the number of inspirals as a function of cluster mass.  In \cite{Rodriguez2015a}, this was done with a simple linear regression, assuming that the mass of the cluster at 12 Gyr was proportional to the number of inspirals it had produced.  However, as we have shown, this scaling is somewhat more complicated: while the mass of a cluster may be proportional to the number of BBHs it produces, the fraction of those sources that will merge in a Hubble time, $f_{\text{insp}}$, is also controlled by the cluster mass.  While the number of BBHs a cluster forms should linearly depend on its mass (more stars yield more BHs, yielding more BBHs), the relationship between $f_{\text{insp}}$ and the final GC mass is less obvious.

Rather than derive a physically-motivated functional form for $f_{\text{insp}}$, we choose a simpler approach, and require only that whatever $f_{\text{insp}}$ we choose increases with mass and assymptotes to 1 as $M_{GC} \rightarrow \infty$.  As this requirement also describes the family of cumulative probability distributions, we elect to use an error function (the cumulative normal distribution).  This yields a final relationship for the number of inspirals as a function of GC mass:

\begin{equation}
N_{\text{insp}}(M_{GC}) =  N_{\text{BBH}}(M_{GC})\times f_{\text{insp}}(M_{GC})
\label{eqn:ninsp}
\end{equation}

\noindent where

\begin{align}
N_{\text{BBH}}(M_{GC}) &= a M_{GC} + b \\ 
f_{\text{insp}}(M_{GC}) &= \erf\left( \frac{M_{GC} - M_{0}}{2\sigma} \right)
\end{align}

\noindent We fit $N_{\text{BBH}}$ and $f_{\text{insp}}$ separately, then take
the product to be $N_{\text{insp}}$.   We fit equation \eqref{eqn:ninsp} for
high-metallicity and low-metallicity clusters.  We also consider separate fits
to clusters with $R_v=1\,\rm{pc}$ and $R_v=2\,\rm{pc}$ to better understand the impact of cluster size on our results (Figure \ref{fig:mergerPerMass}). 

\subsection{Sampling our Inspiral distribution}
\label{subsec:sampling}

In order to create a representative population of BBHs from GCs, we must compare our models to observations of GC systems.  In \cite{Rodriguez2015a}, this was accomplished by integrating  $N_{\text{insp}}(M_{GC})$ over the GC mass function (GCMF), which we assumed to be a log-normal distribution with mean $\log_{10}(M_0) = 5.54$ and width $\sigma_M = 0.52$, based on recent observations of the GC luminosity function in \cite{Harris2014} and an assumed mass-to-light ratio of 2 in solar units \cite{Bell2003}.  This, when combined with the spatial density of GCs per comoving volume \cite[][Supplemental Materials]{Rodriguez2015a}, yielded an effective estimate for the density of BBH mergers from GCs in the universe.  This was then converted into a detection rate by multiplying this density by the distribution of inspirals from our models in redshift, reweighed to favor GC models which more closely resembled the MWGC distribution.  This statistical machinery was used to account for the fact that the models developed in \cite{Morscher2015} and used in \cite{Rodriguez2015a} were designed to probe the space of GC initial conditions, \emph{not} to span the space of observed GCs.

We proceed in a similar fashion.  We integrate equation \eqref{eqn:ninsp} over
the GCMF from 0 to $2\times10^7 M_{\odot}$ to determine the mean number of
inspirals per GC.  Note that we do not consider a varying upper-mass cutoff in
the integral, as the upper-mass cutoff does not strongly affect the mean
\cite[][and its associated erratum]{Rodriguez2015a}.  We do this separately for
high- and low-metallicity clusters, though we use the same GCMF for both.  We
present the results in Table \ref{tab:ninsp}.  To combine the different
metallicity results, we use the computed spatial density of high-metallicity
($\rho_{\text{GC}}^{\text{high}} = 0.34 \,\text{Mpc}^{-3}$) and low-metallicity
clusters ($\rho_{\text{GC}}^{\text{low}} = 0.44 \,\text{Mpc}^{-3}$) from the Supplemental Materials of \cite{Rodriguez2015a}, which implies that 56\% of all clusters are low-metallicity. 

\begin{table}[tb]
\begin{tabular}{l|ccc}
\hline\hline
&& \textbf{Virial Radius} &\\\hline
\textbf{Metallicity}& $R_v = 1\,\rm{pc}$ & $R_v = 2\,\rm{pc}$ & $R_v = 1,2\,\rm{pc}$ \\ \hline
Low                  & 319                            & 194 & 251                           \\
High                 & 381                            & 190 & 271         \\
Both				 & 347                            & 192 & 260           \\\hline\hline
\end{tabular}
\caption{The mean number of inspirals per GC over 12 Gyr, found by integrating
equation \eqref{eqn:ninsp} over the GCMF.  We show separately the average
numbers for clusters with $R_v = 1\,\rm{pc}$ and $R_v = 2\,\rm{pc}$, and the combined fit to both.  We also separate the results according to metallicity.}
\label{tab:ninsp}
\end{table}

In addition to $\left< N_{\text{insp}}\right >$, we need to weigh the models to more closely represent the observed mass distribution of GCs.  To accomplish this, we compute the final mass for our $N=2\times10^5,5\times10^5,10^6$, and $2\times10^6$ models.  We then use the average mass for each set of simulations to divide the GCMF into discrete bins, such that each set of models sits in the center of its respective mass bin.  Each model is then assigned a weight corresponding to the total integrated number of observed GCs in that mass bin.  This is done separately for high- and low-metallicity clusters, with 4 bins for the low-metallicity cluster and 3 for high-metallicity clusters\footnote{We do not use the $2\times10^5$ high-metallicity models, as these disrupted before 12 Gyr, and we are interested only in the population of clusters which survives to the present day.}.  In addition to the contribution from clusters with different masses, we have also assumed that 56\% of clusters in the universe have low metallicities.  However, $2/3$ of our models have low metallicity,  so we multiply the weights of the low-metallicity clusters by 0.86, to ensure our inspirals are correctly sampled according to the metallicity spread of observed GCs.

We use these weights to create a collection of BBH mergers representative of a full population of GCs.  This is accomplished by randomly drawing samples from each model according to its weight.  As an example, the high-metallicity $N=2\times 10^6$ model has the largest weight of all our models, so we draw all of its BBH inspirals for our merger population.  On the other hand, the low-metallicity $N=5\times 10^5$ model has a weight 0.46 times the weight of the largest model, so we only draw 46\% of its binaries.

For the next section, we will use this collection of inspirals for our primary analysis.  For any scatter plots showing a representative sample of mergers, we only use a single draw from our models.  For percentiles, rate computations, and any quantities involving relative quantities, we perform 10 independent draws from our GC models.  This effectively considers each BBH merger from each of our models, weighted by the contribution of that model to the total merger rate.

\section{BBH Properties}
\label{sec:bbhprop}

\subsection{Mass Distributions}

\begin{figure}[tb]
\centering
\includegraphics[trim=3cm 0cm 3cm 0cm,scale=0.75]{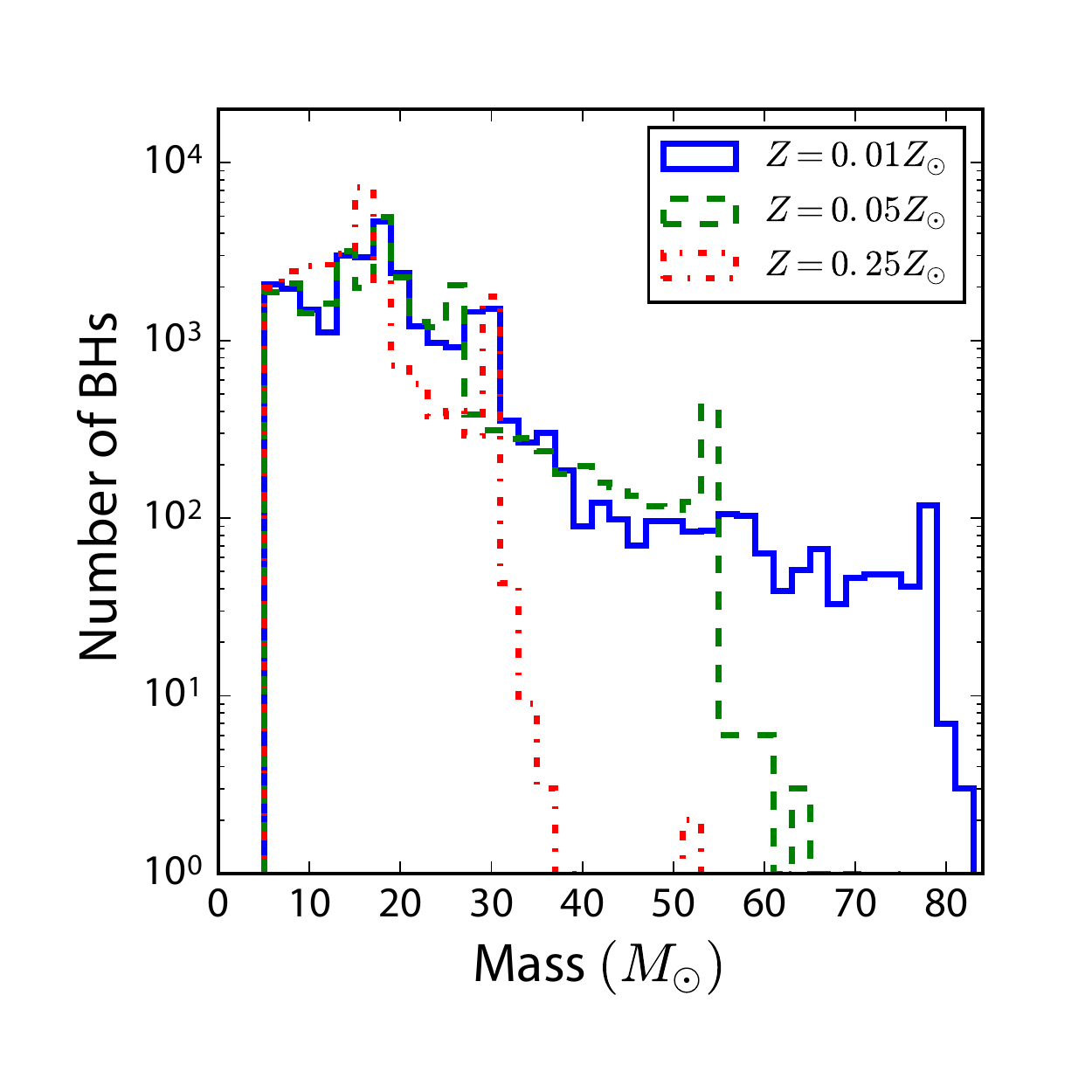}
\caption{The mass distribution of single BHs from our models as a function of metallicity.  As the metallicitiy decreases, the amount of mass loss due to stellar winds is also decreased, increasing the maximum BH mass.  For higher metallicity systems, the wind-driven mass loss causes all stars above a certain mass to form BHs with similar masses (e.g.,\ $30M_{\odot}$ for stars with $0.25Z_{\odot}$).}
\label{fig:massspectrum}
\end{figure}

\begin{figure*}[tb!]
\centering
\includegraphics[trim=3cm 0cm 3cm 0cm,scale=0.6]{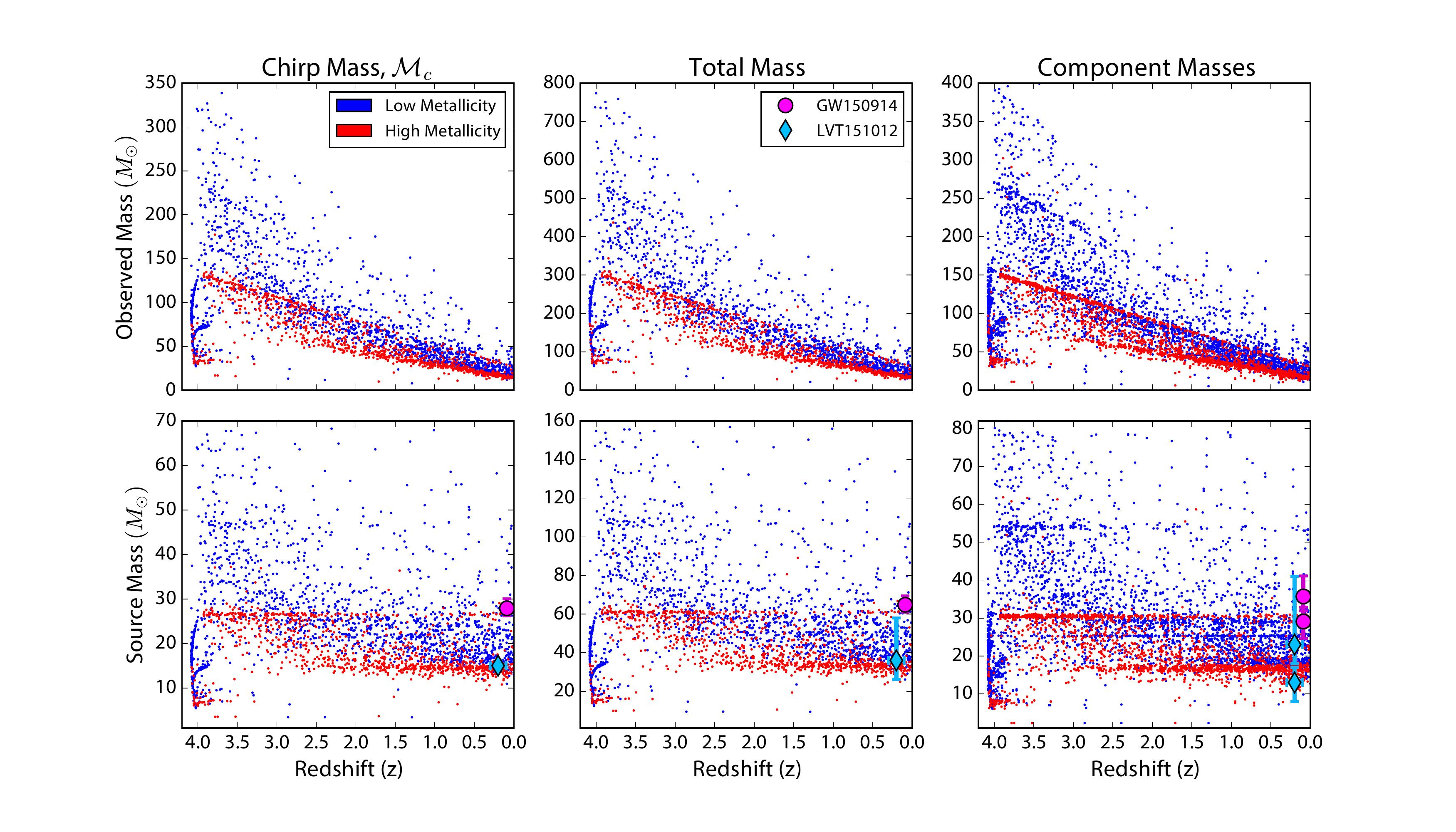}
\caption{Scatter plot of BBH merger masses, weighted to select more inspirals from models with final GC masses near the peak of the GCMF (see Sec.\ \ref{subsec:sampling}).  We show separately the results from models with $Z = 0.25Z_{\odot}$ (in red) and $Z = 0.05,0.01Z_{\odot}$ (in blue).  Along the top, we show the chirp mass, total mass, and individual component masses for binary mergers as observed in the detector frame (i.e. $m_z = m(1+z)$), while the bottom shows the intrinsic masses as measured at the source.  Note that the plot range excludes 5 sources at very high masses (total mass $\sim 250M_{\odot}$) from the chirp mass and total mass plots, and 18 points from the component-mass plot, which are the result of repeated mergers of BH progenitors early in the GC evolution.  We also show the source-frame masses of GW150914 (in magenta) and the GW trigger LVT151012 (in teal), with the 90\% intervals reported from the GW parameter estimation \cite{LSCPE2016,LSCBBHSearch2016}.  Although it was not claimed as a detection, LVT151012 has a $\gtrsim 84\%$ probability of having an astrophysical origin \cite{Abbott2016b}.  Due to the lack of published uncertainties, the LVT151012 total mass intervals are computed by adding the 90\% credible intervals on the individual components from \cite{LSCBBHSearch2016}.}
\label{fig:masses}
\end{figure*}

\begin{figure*}[tb!]
\centering
\includegraphics[trim=3cm 0cm 3cm 0cm,scale=0.6]{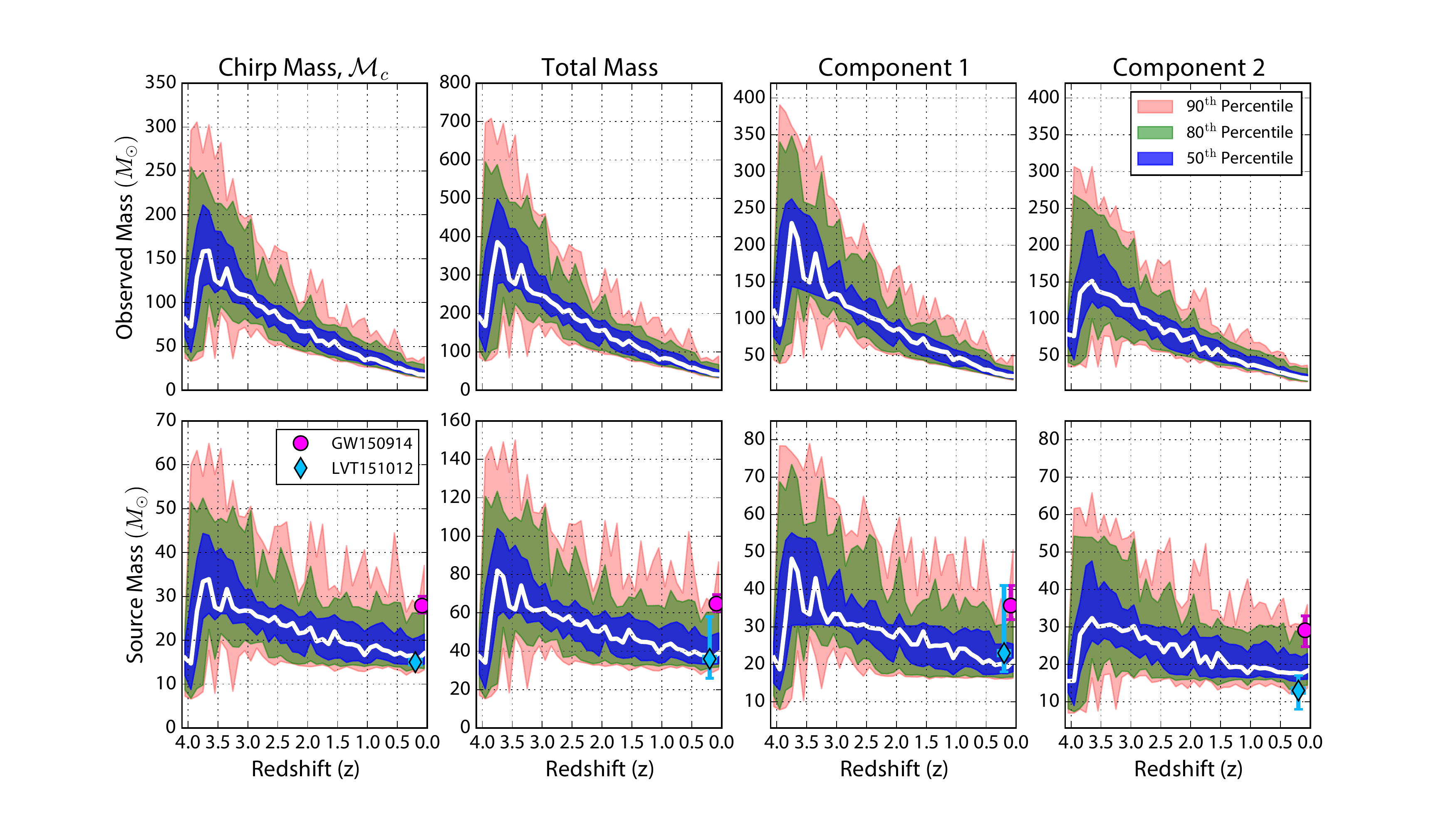}
\caption{Percentage of sources in a given mass range as a function of redshift,
computed from a sample containing 10 weighted draws of inspirals from our
models.  The sources are divided into bins with width $\Delta z = 0.1$.  The
solid white line represents the median mass at a given redshift, while the blue,
green, and red intervals show the mass range containing 50\%, 80\%, and 99\% of
all inspirals at that redshift.  We separate the two component masses, with
component 1 greater than component 2.  We also show GW150914 and
LVT151012, in magenta and teal respectively.  The component BH masses of
GW150914 are consistent with the high-mass end of the BBH mass distribution (the
80\% region), while LVT151014 lies nearly at the median total mass of merging
BBHs.}
\label{fig:massbins}
\end{figure*}

With the Vink prescription for stellar winds, our models now produce significantly more massive BHs than was found in previous GC studies.  While \cite{Rodriguez2015a} found a maximum BH mass of $\sim20M_{\odot}$ and $\sim 25M_{\odot}$ for high and low-metallicity systems respectively, the reduced stellar winds for low-metallicity massive stars significantly increase these cutoffs.  We now find maximum BH masses of $\sim30M_{\odot}$, $\sim53M_{\odot}$, and $\sim78M_{\odot}$ for our $0.25 Z_{\odot}$, $0.05 Z_{\odot}$, and $0.01 Z_{\odot}$ models respectively (see Figure \ref{fig:massspectrum}).  This is consistent with the latest results from population synthesis codes, such as \texttt{StarTrack} \cite{Belczynski2010a} and \texttt{SEVN} \cite{Spera2015a} (though the latter, which explores significantly higher-mass progenitors than we consider here, can produce BHs as massive as $130 M_{\odot}$ for stars with $Z = 0.01 Z_{\odot}$).

Despite the significant changes to the BH mass spectrum, the behavior of BHs in clusters, developed in \cite{Morscher2015}, remains unchanged: after core collapse, the most massive BHs segregate into the center of the cluster, where they immediately form BBHs via three-body encounters.  These BBHs then undergo a series of binary-single and binary-binary encounters, increasing their binding energies  and shrinking their semi-major axes.  Eventually, the recoil from one of these encounters will be sufficient to eject the binary from the cluster, as discussed in Section \ref{subsec:binprop}.  Although a significant number of binaries merge in the cluster ($\sim 10\%$), the majority of these in-cluster inspirals occur early in the GC lifetime.  At $z < 1$, only 0.06\% of binary mergers (one merger from all 48 models) occur in-cluster.  Of the ejected sources merging in the local universe, 99.7\% were formed dynamically, which we define to be either a BBH formed from three isolated BHs by a three-body interaction, or a BBH formed from a primordial binary which swapped components at least once during a binary-single or binary-binary encounter.

In Figure \ref{fig:masses}, we show the masses for each of the inspirals from the weighted sample of GC BBH mergers.  We break the masses down into two categories: source masses, or the local masses of each BBH, and observed masses, which correspond to the redshifted mass, $m_z = m(1+z)$, measured by an observer on Earth.  We also show separate panels for the chirp mass of the source, $\mathcal{M}_c \equiv (m_1m_2)^{3/5}/(m_1 + m_2)^{1/5}$, the total mass of the source, and the individual components of each binary.

The overall structure of the plots agrees well with our understanding of BH and BBH evolution in GCs: after the formation and core collapse of the cluster (at $z\sim 4$), the most massive BHs form binaries and are ejected immediately.  The GC processes through its available population of BHs, working its way through the BH population from most to least massive, so that only low-mass BHs ($\sim 10-20 M_{\odot}$) are still present in massive GCs by the present day.  In the total-mass panel of Figure \ref{fig:masses}, this story is obvious.  The majority of the most massive inspirals (total mass $> 100M_{\odot}$)  merge soon after GC formation, between $z=4$ and $z=3$.  After these BBHs are ejected, the cluster moves on to less massive BBHs with longer inspiral times.  These sources (with total masses from $30M_{\odot}$ to $60M_{\odot}$) form the predominant population of BBHs detectable in the present day.  

The plateaus in the chirp mass, total mass, and component mass distributions  are primarily determined by the maximum BH mass at each metallicity, which is in turn determined by the wind-driven mass loss from the Vink prescription (Figure \ref{fig:massspectrum}).  For the highest metallicity models ($Z = 0.25Z_{\odot}$), this yields a large population of $30M_{\odot}$ BHs, which in turn forms a large population of equal-mass BBHs with a total mass of $60M_{\odot}$.  For the $Z = 0.05Z_{\odot}$ clusters, this yields a smaller collection of sources with a total mass of $110M_{\odot}$.  However, for the lowest-metallicity models ($Z = 0.01Z_{\odot}$) there is no apparent collection of sources at $160M_{\odot}$ as might be expected.  

This behavior can again be explained by the wind-driven mass loss.  Each model begins with an identical distribution of stars drawn from \eqref{eqn:kroupa}.  For the highest-metallicity models, the mass-loss from these winds brings all stars with birth masses from $\sim 80M_{\odot}$ to $\sim 150M_{\odot}$ down to a final progenitor mass of $\sim 30 M_{\odot}$ to $\sim 35 M_{\odot}$ before the supernova occurs.  Essentially, this truncates a large section of the high-mass end of the IMF to a single BH mass; however, for lower-metallicity models, the decreased efficiency of the stellar winds means a lower number of high-mass stars are being converted into maximum-mass BHs, essentially spreading out the high-mass stars over a wider range of BH masses.  The number of maximum-mass BHs between each of our models decreases by roughly a factor of 5 between each of our metallicitiy bins.  This also yields a higher number of inspirals with unequal-mass components for lower-metallicity models, which we discuss in the next section.

We also show in Figure \ref{fig:masses} the masses and 90\% uncertainties
associated with the recent detection of GW150914.  The masses reported from the
parameter estimation \cite{LSCPE2016} of GW150914 are consistent with the masses
of BBH mergers from GCs in the local universe, with the specific masses
($36M_{\odot}$ and $29M_{\odot}$) being easily formed in lower-metallicity GCs.
We also show the less-significant GW trigger LVT151012, with the 90\%
uncertainties taken from \cite{LSCBBHSearch2016}.  Although this signal was not
claimed as a detection, we note that the masses of this event (total mass of
$\sim 36M_{\odot}$) is also consistent with BBH mergers from GCs in the local
universe.

In Figure \ref{fig:massbins}, we convert our scatter plot of BBH mergers into percentiles as a function of redshift.  In the local universe ($z \lesssim 0.1$), we find that the median BBH source from a GC has a chirp mass of $17 M_{\odot}$, with 50\% of sources lying in the range $[14.6M_{\odot} - 21.4M_{\odot}]$, 80\% of sources in the range $[13.2M_{\odot} - 27.7M_{\odot}]$, and 90\% of sources in the range $[13.2M_{\odot} - 37.1M_{\odot}]$.  This corresponds to a median total mass of $39.4M_{\odot}$, with a 50\% interval of $[33.6M_{\odot} - 49.6M_{\odot}]$, a 80\% interval of $[31.8M_{\odot} - 63.6M_{\odot}]$, and a 90\% interval of $[30.5M_{\odot} - 86.8M_{\odot}]$.  Given the analysis in Section \ref{subsec:binprop} and Figure \ref{fig:insp_times}, we can expect that massive binaries that merge in the local universe are more likely to be ejected from clusters with lower masses and larger half-mass radii.  This is supported by our results: 77\% of all BBH mergers are formed in our most massive GCs ($N=2\times10^6$). However, in the local universe, this fraction changes as a function of mass.  86\% of binary mergers with total masses $< 50M_{\odot}$ at $z\sim0.1$ are preferentially formed in high-mass clusters.  On the other hand, for binary mergers with masses above $50M_{\odot}$, this fraction drops to 66\%.  We note that the large contribution from high-mass GCs is a result of our focus on GCs which have survived to the present day.  Were we to consider lower mass clusters that disrupted before 12 Gyr, it is likely that the merger rate for massive BBHs in the local universe would increase significantly.

Finally, we note that there exists a small population of BHs significantly more massive than the maximum-BH mass allowed by stellar evolution.  These $\sim 25$ BHs are the result of repeated stellar mergers early in the evolutionary history of the cluster.  These collisions are primarily the mergers of massive giant and main-sequence stars during binary-single and binary-binary encounters, with a smaller number resulting from direct collisions of single stars.  We note that 3 of these sources were formed from a BBH merger whose remnant remained in the cluster, forming a second BBH; however, as CMC does not incorporate the physics of gravitational-wave recoil, these 3 sources are most likely nonphysical, as typical recoil velocities from a BBH merger greatly exceed the escape velocity of a typical GC \cite{Campanelli2007}.

\subsection{Mass Ratio Distributions}

\begin{figure}[tb]
\centering
\includegraphics[trim=1cm 0cm 1cm 0cm,scale=0.57]{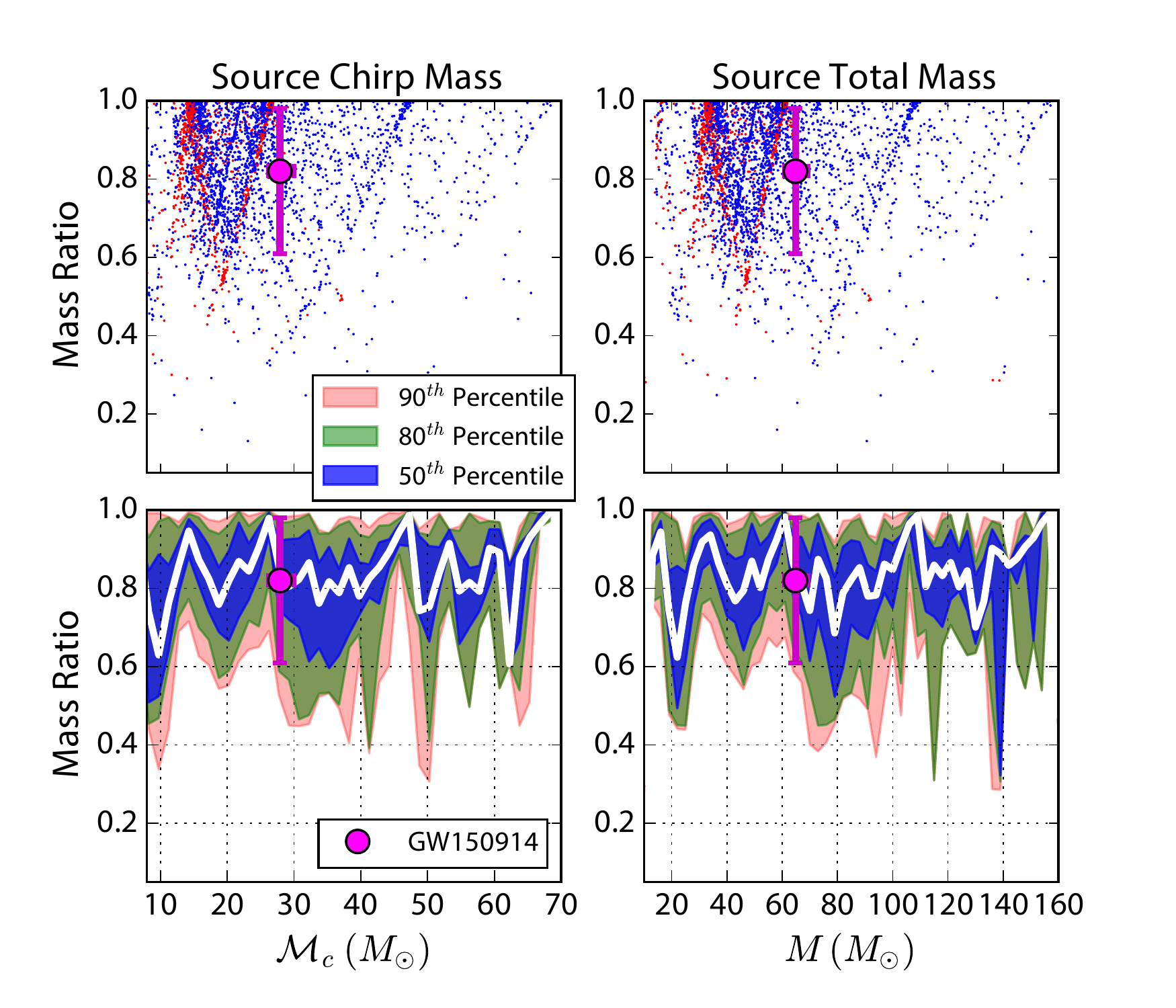}
\caption{The mass ratios for all inspirals from Figures \ref{fig:masses} and
\ref{fig:massbins}.  On the top, we show the scatter plot of mergers, with red
and blue indicating inspirals from high and low-metallicity GCs, respectively.
The bottom panels indicate the median mass ratio for a given source mass, with
bin widths of $\Delta \mathcal{M}_c = 1.5M_{\odot}$ and $\Delta M_{\text{tot}} =
3M_{\odot}$, while the blue, green, and red ticks indicate the range of mass
ratios encompassing 50\%, 80\%, and 90\% of all sources.  We also show
the mass ratio for GW150914 and its 90\% credible region.  We do not show the
LVT151012 trigger, which lacks a published mass ratio.}
\label{fig:ratios}
\end{figure}

In addition to the individual BH masses, the mass ratios of BBHs may provide an important clue to the formation mechanism of a binary.  In Figure \ref{fig:ratios}, we show the mass ratios from our models as a function of chirp mass and total mass.  The mass ratios pile up near unity, with a median mass ratio of 0.87 and 68\% of sources having mass ratios greater than 0.8.  This is to be expected: binary-single and binary-binary scattering experiments show that binaries in dense stellar environments are prone to swapping components, preferentially ejected less-massive components in favor of more massive companions \cite{Sigurdsson1993}.

However, we also note that there exists a small population ($7\%$) of sources with mass ratios less than 0.6.  These appear at the base of the apparent ``V'' formations in the top panels of Figure \ref{fig:ratios}, corresponding to binaries with total masses of $\sim 20M_{\odot}$, $\sim 50M_{\odot}$, $\sim 80M_{\odot}$, and $\sim 130 M_{\odot}$.  These features are a direct result of the peaks in the distribution of BH masses (Figure \ref{fig:massspectrum}).  The three most-obvious peaks in the BH mass distribution, around $18M_{\odot}$, $30M_{\odot}$, and $53M_{\odot}$, create peaks of equal-mass binaries at total masses near $36M_{\odot}$, $60M_{\odot}$, and $106M_{\odot}$.   For instance, the $\sim 53M_{\odot}$ maximum BH mass for the $0.05Z_{\odot}$ models creates a large population of equal-mass binaries with a total mass of $106M_{\odot}$.  However, these BHs will also form binaries with less-massive companions from the BH distribution (roughly down to the second peak of the distribution at $\sim 25M_{\odot}$).  This creates the feature running from a mass ratio of 1 at $\sim 106 M_{\odot}$
 to a mass ratio of 0.4 at $\sim 80 M_{\odot}$. We reiterate that these features are the result of the peaks in the BH mass distribution, which strongly depends on the stellar metallicity.  As we have only considered three metallicities here, it is likely that a collection of models fully spanning the distribution of GC metallicities would find a more continuous distribution of mass ratios.  
 
\subsection{Eccentricity Distributions}

It is a well-known result that dynamically-formed binaries will follow a thermal
distribution of eccentricities \cite{Heggie1975}.  This
distribution, $P(e)de = 2ede$, is the result of the thermalization of velocities
that occurs through repeated encounters in the dense cluster core.
However, once ejected, these binaries evolve in isolation, and are rapidly
circularized due to the preferential emission of gravitational radiation at
periapsis \cite{Peters1964}.  Given the difficulties involved in
detecting such binaries and the possibility of parameter
estimation for such systems \cite{Huerta2014}, it is important to quantify the number of
sources that will enter the LIGO detection band with non-negligible
eccentricities.
 
\begin{figure}[tb]
\centering
\includegraphics[trim=3cm 2cm 3cm 2cm,scale=0.6]{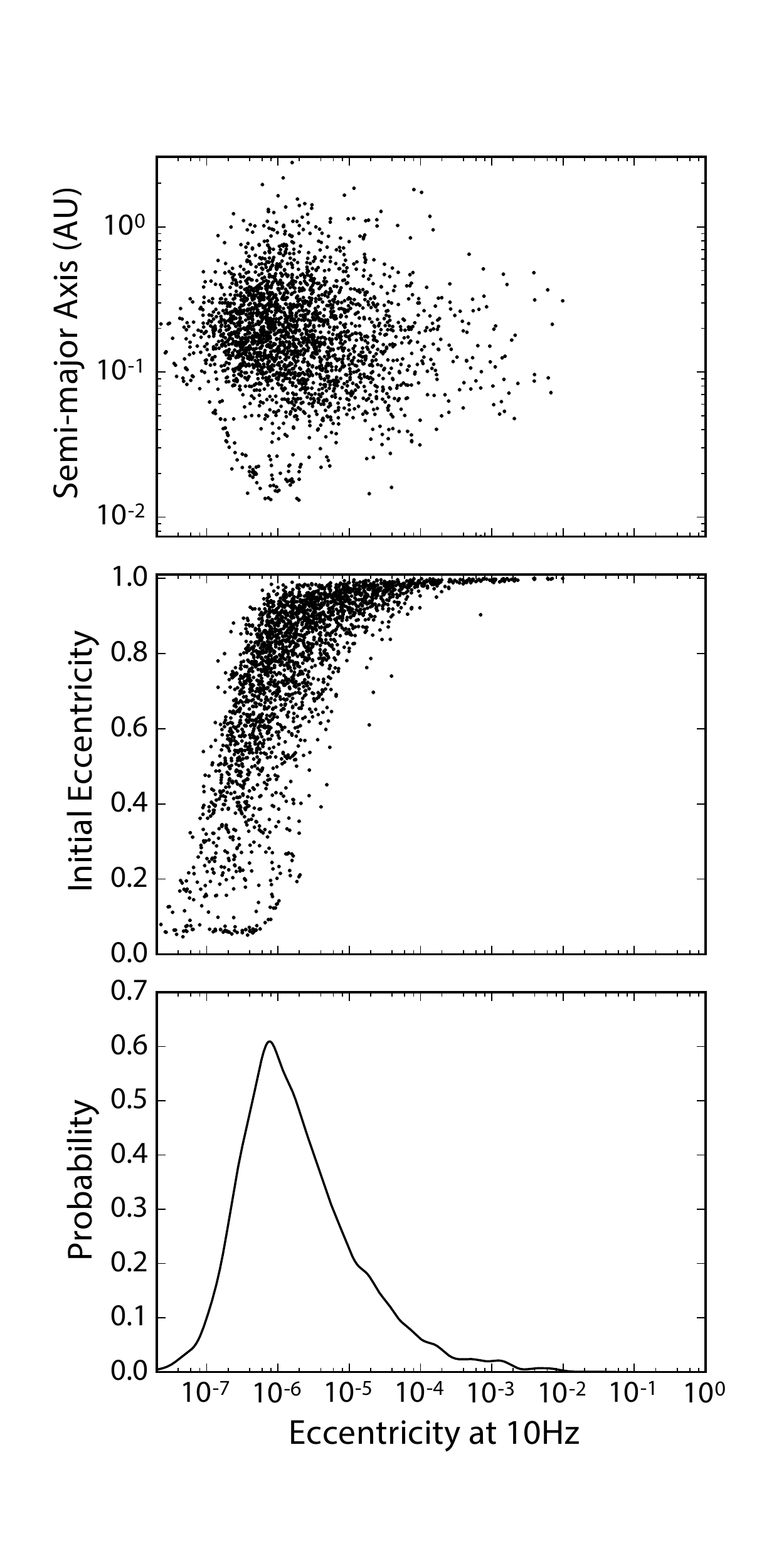}
\caption{The eccentricities of the BBH population at a gravitational-wave frequency of 10Hz.  The top plot shows the final eccentricities as a function of the semi-major axis at ejection (which corresponds to 99.9\% of sources that merge at $z<1$).  The middle plot shows the eccentricity at ejection vs. 10Hz.  The bottom plot shows the distribution of all sources at 10Hz.}
\label{fig:ecc}
\end{figure}

 In Figure \ref{fig:ecc}, we show the eccentricities of all ejected binaries.
 The eccentricity at 10Hz is computed by integrating $\left<de/da\right>$ \cite{Peters1964} from ejection to $a_f = G(m_1 + m_2) / (4\pi^2f_{\text{low}}^2)$, the Keplerian
 separation for a binary with an orbital period of $f_{\text{low}} = 5$Hz
 (corresponding to a gravitational-wave frequency of 10Hz).  We show the
 relationship between the 10Hz eccentricity and the semi-major axis and
 eccentricity at ejection in the upper and middle plots, and the distribution of
 final eccentricities in the lower plot.  

 This result suggests that the majority of BBHs from GCs will very nearly circular by the time they are detectable, with 99\% of sources
 entering the LIGO band with eccentricities below $\sim10^{-3}$.  Thus, eccentric
 systems ejected from clusters will not be a significant source for Advanced
 LIGO.  However, it has been shown that a non-negligible number of sources in the cluster
 will be driven to merger by long-term secular effects, such as the Lidov-Kozai mechanism, in BH triple
 systems, which we do not consider here.  Were that physics
 properly included in these models, it is likely that $\sim$1\% of
 mergers could be detected with significant eccentricity \cite{Antonini2015}.

\subsection{The Impact of Stellar Evolution}
\label{subsec:se}

\begin{figure}[tb]
\centering
\includegraphics[trim=3cm 0cm 3cm 0cm,scale=0.75]{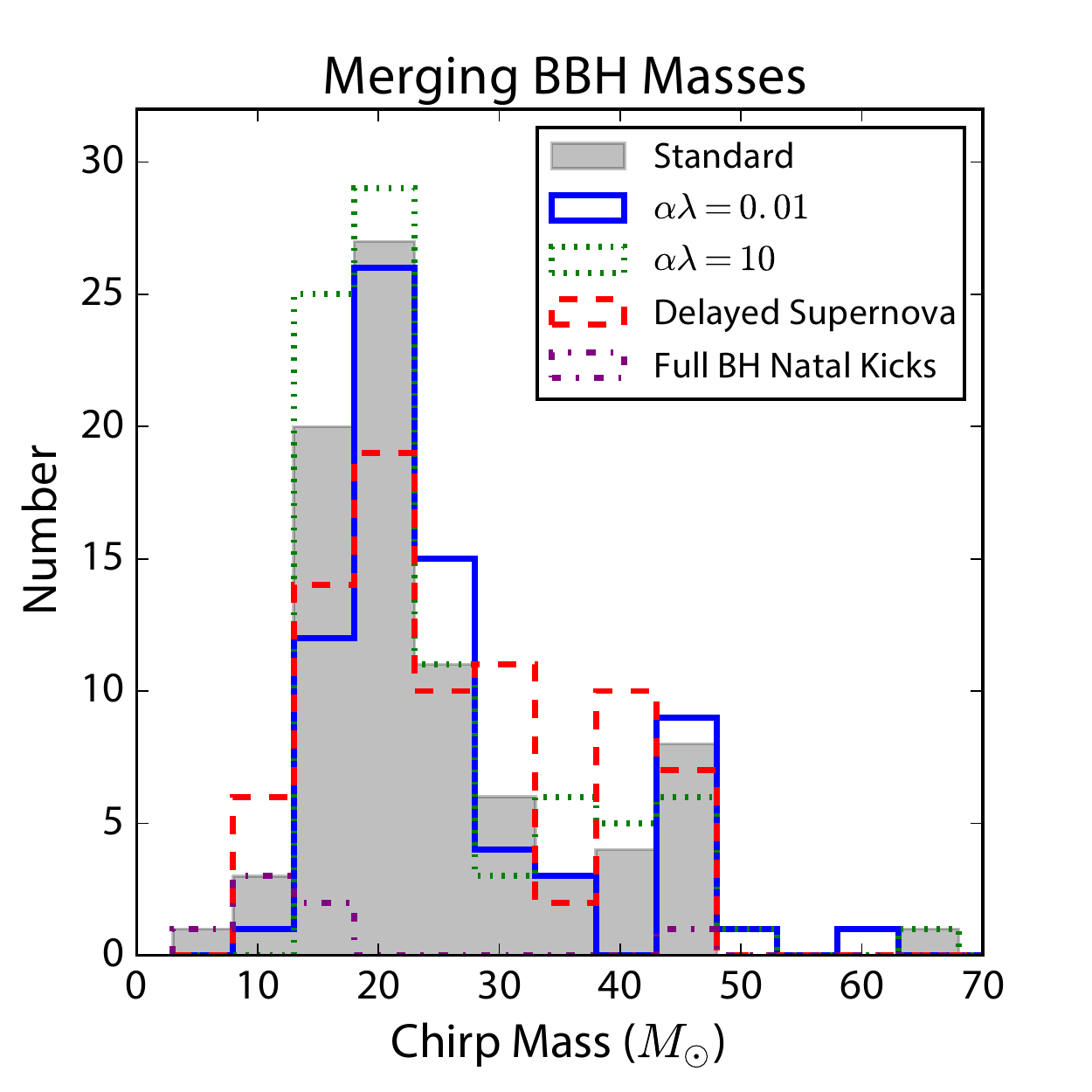}
\caption{Chirp masses of merging binaries from the $N=10^6$, $Z=0.05Z_{\odot}$,
$R_v=1\,\rm{pc}$ model.  We compare our standard model to models with different values of $\alpha\lambda$ for the common envelope, a model with a different SN prescription, and a model where BHs are born with natal kicks equivalent to NSs.  All changes to the binary stellar evolution produce no significant difference.  However, the model with high natal kicks significantly reduces the number of BBH inspirals.}
\label{fig:se}
\end{figure}

The main uncertainty in population synthesis estimates of the BBH merger rate from the field concerns the physics of binary stellar evolution.  The most recent studies \cite[e.g.,][]{Dominik2012} have considered a large number of possible outcomes from various stages of binary stellar evolution.  These different models are then used to constrain the uncertainties, although the sheer number of parameters makes it difficult to constrain the merger rate even to within several orders-of-magnitude.

More recent work \cite{Belczynski2015} has focused on three specific questions: can BBH progenitors survive the common-envelope (CE) phase when the donor star is in the Hertzsprung gap (HG)?  What is the standard mechanism for BH-producing supernova (rapid or delayed)?  And what is the impact on the rate if BHs are born with natal kicks drawn from the same distribution as NSs?  The standard \texttt{StarTrack} model assumes that stars in the HG cannot be CE donors, that supernovae are well described by the rapid model, and that BH natal kicks are reduced proportionally to the mass of fallback material, as in equation \eqref{eqn:kick}.  

In contrast, our models for dynamical formation of BBHs are largely insensitive to the uncertainties in binary stellar evolution.  Since the majority of BBHs are formed dynamically, their numbers, semi-major axes, eccentricities, and inspiral times are  determined by well-understood gravitational physics.  Any changes to our binary stellar evolution prescription should not have a significant effect on our BBH population.  However, the same cannot be said for single star evolution: any changes that modify the number or masses of single BHs will significantly change both results from both clusters and the field.

To quantify these effects, we rerun our $N=10^6$, $Z=0.05Z_{\odot}$, and $R_v =
1\,\rm{pc}$ model, varying some of the assumptions in binary and single star evolution.  We consider 5 models: 

\begin{itemize}
\item the \textbf{standard model}, adopted in the previous sections.  We use the standard BSE prescription for the CE evolution, employing the $\alpha \lambda$ formalism\footnote{The term $\alpha \lambda$ arises from the energy balance between the gravitational potential energy of the envelope and the orbital energy of the binary: \begin{equation}
\alpha \Delta E_{\text{bin}} = \frac{G M_d M_e}{\lambda R}
\end{equation} where $\Delta E_{\text{bin}}$ is the change in binding energy of the binary before and after the CE, $M_d$ is the mass of the donor star, $M_e$ is the mass of the donor star's envelope, and $R$ is the donor star's radius at Roche lobe overflow.  In this model $\lambda$ parameterizes  the binding energy of the envelope, while  $\alpha$ parameterizes  the efficiency of energy transfer from the binary's shrinking orbit to the envelope.} \cite{Webbink1984}, with $\lambda$ calculated by a fitting formula similar to that described in \cite{Claeys2014}.  By default, BSE will allow binaries to survive the CE phase so long as the core of either star does not fill its Roche lobe (including stars in the HG, for which BSE employs a time-dependent density profile \cite{Hurley2002}). 

\item \textbf{low CE binding energy model}, in which we set $\alpha \lambda = 0.01$ for all binaries,

\item \textbf{high CE binding energy model}, in which we set $\alpha \lambda = 10$, 

\item \textbf {delayed supernova model}, in which we use the delayed supernova model for BH masses and fallback, as described in \cite{Fryer2012}, and

\item \textbf{full BH natal kicks model}, in which we ignore the effects of fallback, and give each BH its full kick velocity from the NS distribution.
\end{itemize}

We show the impact of these different cases on the distribution of BBH mergers in Figure \ref{fig:se}.  As expected, the mass spectrum of BHs does not significantly depend on the physics of stellar evolution, with one notable exception for the full BH natal kicks case.  Over 12 Gyr of evolution, the standard model produces 84 mergers.  The $\alpha\lambda = 0.01$ case produces 72 mergers, the $\alpha\lambda = 10$ case produces 87 mergers, the delayed supernova case produces 79 mergers, but the full kicks case produces only 7 mergers.  

This significant decrease in BBH production is to be expected: the distribution of BH velocities in the high-kick case is drawn from a Maxwellian with $\sigma=265$ km s$^{-1}$, meaning most BHs will be born with speeds significantly greater than the escape speed from the center of the cluster ($\sim50-100$ km s$^{-1}$).  In that case, the majority of BHs are ejected before they have a chance to dynamically form binaries.  We explore the implications of this for the BBH merger rate in Section \ref{subsec:field}.

\section{Cosmological Merger Rates}
\label{sec:rates}

\begin{figure*}[tb]
\centering
\includegraphics[trim=3cm 0cm 3cm 0cm,scale=0.6]{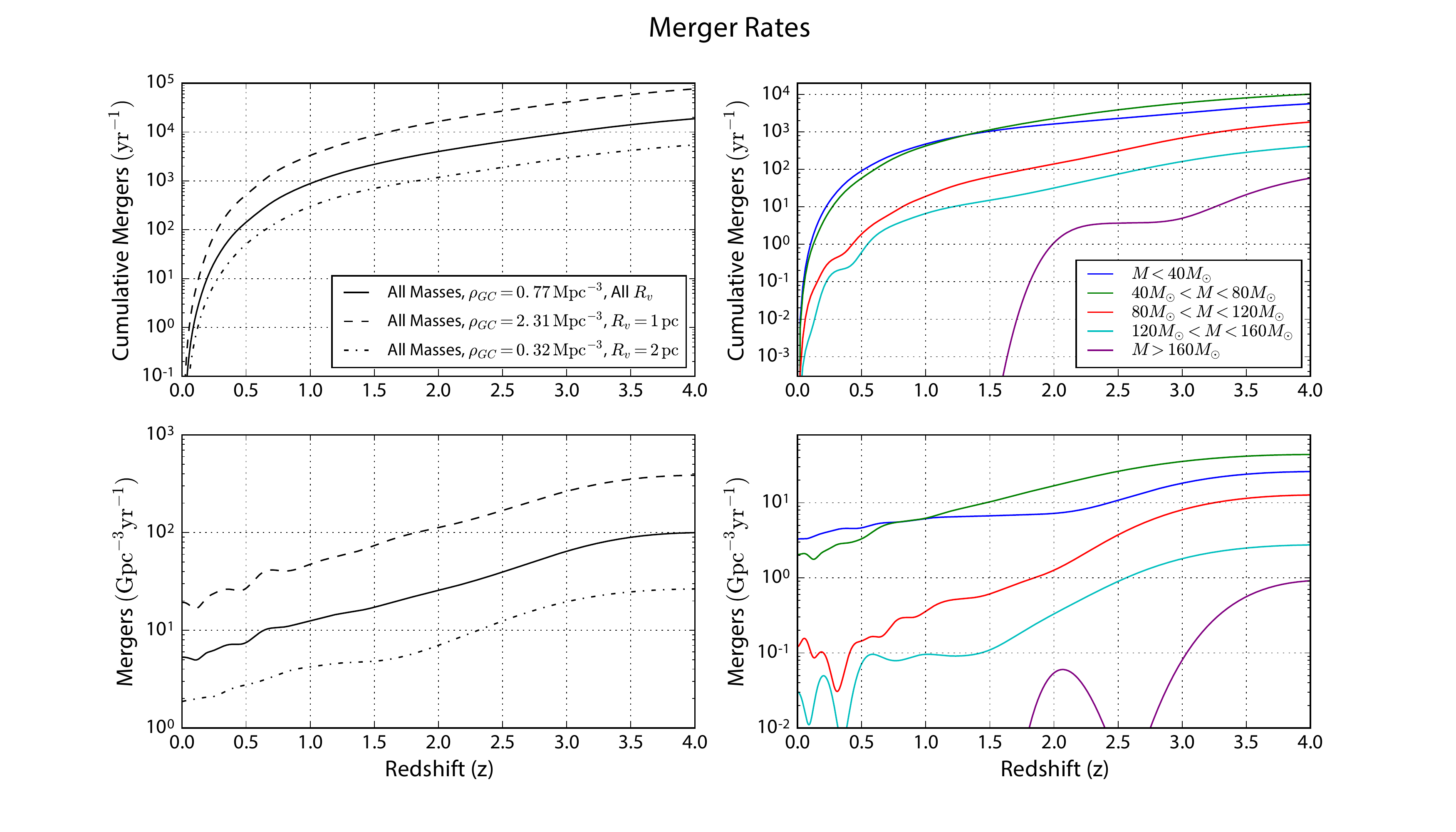}
\caption{The BBH merger rates from our models as a function of redshift.  The upper panels show the cumulative rate of mergers per year in a volume out to redshift $z$, with the left panel showing the cumulative merger rate for all binaries, and the right panel showing the cumulative merger rate for binaries with specific total masses.  The lower panels show the source merger rate in Gpc$^{-3}$yr$^{-1}$ at a given redshift for all BBHs (left) and for specific BBH total masses (right).  For the total merger rates (the leftmost panels) we illustrate the uncertainties in our models to specific assumptions, showing how the rate varies with the spatial density of GCs and our choice of initial virial radius.}
\label{fig:rates}
\end{figure*}

To compare different models for BBH formation, it is necessary to understand the predicted merger rate for each different formation scenario.  We compute the cosmological merger rate of BBH inspirals from our models as a function of redshift.  We then assume all GCs to be 12 Gyr old, and we describe the merger rate per unit time and comoving volume as 

\begin{equation}
\mathcal{R}_s(z) = \rho_{GC} \left< N_{\text{insp}} \right> P(z)
\label{eqn:sourcerate}
\end{equation}

\noindent where $\rho_{GC}$ is the spatial density of GCs \cite[][Supplemental Materials]{Rodriguez2015a}, $\left< N_{\text{insp}} \right>$ is the mean number of GC inspirals per cluster from Section \ref{subsec:mergerPerClus}, and $P(z)$ is the normalized merger rate at a given redshift, defined as

\begin{equation}
P(z) = P_t(t_{\text{lookback}}(z))
\label{eqn:pz}
\end{equation}

\noindent where $P_t$ is the probability distribution of inspiral merger times, computed by generating a kernel density estimate of the merger times from our sample of BBH mergers, and  $t_{\text{lookback}}$ is the cosmological lookback time at a given redshift \cite{Hogg1999a}.  Note that $P(z)$ is the distribution of sources \emph{in time} at redshift $z$, not the distribution of sources in redshift\footnote{By computing the kernel density estimate with the inspiral times of the bianries, the probability of seeing an inspiral at time $t$ is $P_t(t)dt$.  By converting this to the distribution in redshift with \eqref{eqn:pz}, we ensure that $P(z)$ is in units of equal time, not equal redshift.  This is different than the approach employed in \cite{Rodriguez2015a}, and we note that properly accounting for the distribution increases the detection rates from that study by $\sim 30\%$.}.

Additionally, we are interested in the total number of sources that merge within a comoving volume out to a given redshift, $z$.  We compute the observed rate of mergers per unit time as

\begin{equation}
\mathcal{R}_o(z) = \int^{z}_0 \frac{dV_c}{dz'} \mathcal{R}_s(z') \left(\frac{dt_s}{dt_o} \right) dz'
\label{eqn:obsrate}
\end{equation}

\noindent where $dV_c/dz$ is the comoving volume at a given redshift, $\mathcal{R}_s(z')$ is the source merger rate \eqref{eqn:sourcerate}, and $dt_s/dt_o=(1+z)^{-1}$ is rate of time dilation between a clock measuring the merger rate at redshift $z'$ and a clock on Earth \cite[e.g.,][]{Belczynski2014}.   

To gain a handle on the uncertainties in our assumptions, we consider three
different cases: our standard case assumes a spatial density $\rho_{GC}=0.77$
Mpc$^{-3}$, and that half of those clusters have virial radii of $1\,\rm{pc}$, and half
have virial radii of $2\,\rm{pc}$.  We also consider an optimistic case, which assumes
$\rho_{GC}=2.31$ Mpc$^{-3}$ (the optimistic estimate from \cite{Rodriguez2015a})
and $R_v = 1\,\rm{pc}$ for all clusters, and a conservative case, which assumes
$\rho_{GC}=0.33$ Mpc$^{-3}$, and $R_v = 2\,\rm{pc}$ for all clusters.  The standard,
optimistic, and conservative cases are shown in the left panels of Figure
\ref{fig:rates}.  The upper-left panel shows the observed cumulative rate as a
function of redshift (equation \ref{eqn:obsrate}), while the lower-left panel
shows the source rate (equation \ref{eqn:sourcerate}).  In the local universe,
our standard assumptions leads to a merger rate of $\sim$5 Gpc$^{-3}$yr$^{-1}$
while our optimistic and conservative assumptions predict merger rates of
$\sim$20 Gpc$^{-3}$yr$^{-1}$ and $\sim$2 Gpc$^{-3}$yr$^{-1}$, respectively.
This is consistent with the merger rate for GW150914-like binaries, which was found to be $2-53$ Gpc$^{-3}\,\text{yr}^{-1}$ \cite{Abbott2016b}.

In the right panels of Figure \ref{fig:rates}, we show the merger rates for binary inspirals with specific total masses from our standard case.  We compute $\left<N_{\text{insp}}\right>$ separately for each mass bin by fitting the inspirals in each bin to equation \eqref{eqn:ninsp} and integrating over the GCMF; this is the same technique used for the total rate, although for mass bins with only a handful of inspirals, such as the 5 inspirals in the $M>160M_{\odot}$ bin, we abandon equation \eqref{eqn:ninsp} in favor of a simple linear regression.  We then compute the distribution of inspirals, $P(z)$, separately for each mass bin, and compute the rates with equations \eqref{eqn:sourcerate} and \eqref{eqn:obsrate}.  In the local universe, we can expect a merger rate of $\sim 3$ Gpc$^{-3}$yr$^{-1}$ for BBHs with a total mass of $M < 40M_{\odot}$, and $\sim 2$ Gpc$^{-3}$yr$^{-1}$ for $40 M_{\odot} < M < 80M_{\odot}$, with higher masses contributing minimally to the total merger rate.  We note that the oscillatory behavior of the high-mass bins at low redshifts is the result of the small number of events.  As an example, the bump at $z=2$ for $M > 160M_{\odot}$ sources is the result of a single inspiral whose BH progenitor underwent repeated mergers.  Given the uncertainties in modeling stellar collisions and mergers, particularly with the Monte Carlo method, the merger rates for such rare events should be treated with appropriate skepticism.  

\subsection{Comparison to the Field}
\label{subsec:field}

\begin{figure*}[htb!]
\centering
\includegraphics[trim=3cm 3cm 3cm 1cm,scale=0.6]{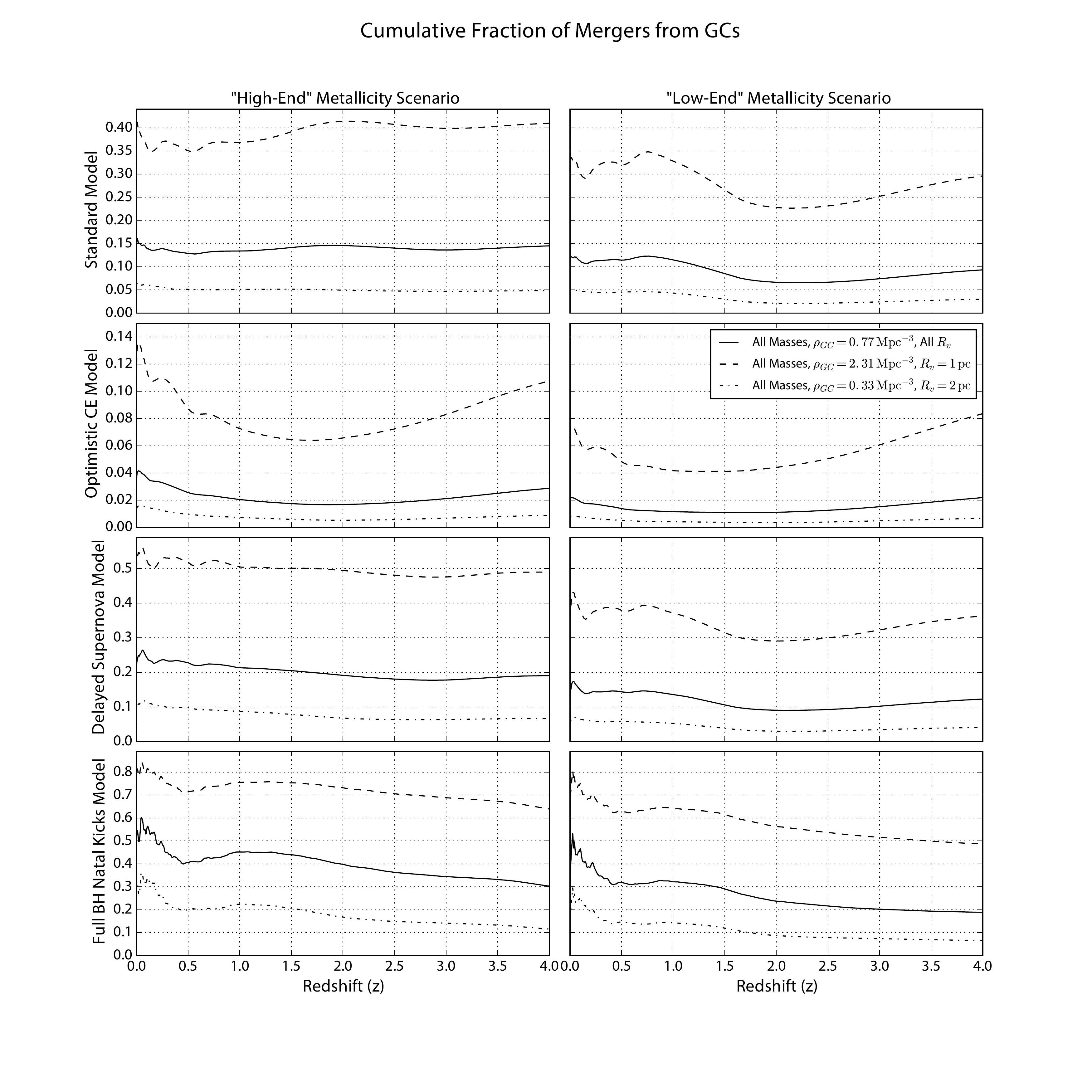}
\caption{The fraction of all BBH mergers that arise from GCs, found by combining the cumulative estimates from Figure \ref{fig:rates} with the merger rate estimates from \cite{Dominik2013} (from the Synthetic Universe website).  We compare the model considered here to each of the stellar evolution and star formation scenarios considered in \cite{Dominik2013}.  We use the same cumulative merger rate for each comparison with the exception of the high BH natal kicks comparison, where we multiply the cumulative merger rate from GCs by $(7/84)$, the fractional decrease in mergers found in Section \ref{subsec:se}.}
\label{fig:fieldrates}
\end{figure*}

One of the basic questions for gravitational-wave astrophysics is whether or not GW observations can discriminate between different formation physics for observed sources.  Can intrinsic parameters, such as the BH masses and spin, be used to determine whether a particular event was formed by dynamics, or by isolated binary stellar evolution?  And can a collection of GW observations be used to answer detailed questions about star formation and stellar evolution \cite{Stevenson2015}?  While each of these questions is worthy of study, in this paper we focus on a much simpler, but more critical question: what fraction of detectable sources originated in GCs?

To that end, we make the following assumptions: we assume that all BBH mergers in the universe arise from either isolated binary evolution in the field \cite[as done in][]{Dominik2012,Dominik2013,Dominik2014} or from GCs.  We also assume that the star formation scenarios for the field and for old GCs are distinct, with all GCs being formed in a single burst of star formation 12 Gyr ago, while star formation in the field occurring continuously, with metallicity increasing with the age of the universe.  The \texttt{StarTrack} models in \cite{Dominik2013} employed a star-formation rate from \cite{Strolger2004} and a metallicity-redshift relationship found by averaging the results of \cite{Pei1999} and \cite{Young2007}; however, these assumptions produced unphysically-high metallicities for star formation in the local universe (with metallicities as high as $3Z_{\odot}$), so they considered two extreme scenarios: a \emph{high-end metallicity scenario}, where they divide their metallicity profile by 1.7, to agree with results from \cite{Yuan2013}, and a \emph{low-end metallicity scenario}, where they divide the profile by 3.0, to agree with SDSS observations from \cite{Panter2008}.  

We compare the cumulative merger rates from our models to the cumulative rates for the \texttt{StarTrack} models examined in \cite{Dominik2013} in Figure \ref{fig:fieldrates}.  The four models considered are: 

\begin{itemize}
\item the \textbf{standard model}, in which they assume all binaries which enter the CE with a donor star in the HG will immediately merge, use the rapid supernova prescription, and employ a physical $\alpha\lambda$ description taken from \cite{Xu2010},
\item the \textbf{optimistic model}, where they allow BBH progenitors to survive the CE with donor stars in the HG,
\item the \textbf{delayed supernova model}, where they use the delayed supernova model, and
\item the \textbf{full BH natal kick model}, where they give newly formed BHs natal kicks identical to those of NSs, regardless of the fallback fraction.
\end{itemize}

\noindent Given that our merger rates are insensitive to the assumptions of binary stellar evolution, we assume that our main results will remain unchanged under the assumptions of the standard, optimistic, and delayed supernova cases.  For the case of high BH natal kicks, we multiply our standard merger rates by 0.08, proportional to the decreased number of mergers from our high-kick $N=10^6$ model (Section \ref{subsec:se}).  This oversimplifies the true relationship between BH retention and natal kicks, since the fraction of retained BHs will increase for more compact, massive clusters with higher escape speeds, but it is sufficient for this current estimate.  We also compare the GC merger rate to the rate from the field for both the high-end and low-end metallicity evolution scenarios.  The fraction of mergers from GCs up to a given redshift $z$ is defined to be 

\begin{equation}
F_{GC}(z) = \frac{\mathcal{R}_o^{\text{GC}}(z)}{\mathcal{R}_o^{\text{GC}}(z) + \mathcal{R}_o^{\text{Field}}(z)}
\label{eqn:cumfrac}
\end{equation}

\noindent where $\mathcal{R}_o^{\text{GC}}(z)$ is from equation \eqref{eqn:obsrate}, and $\mathcal{R}_o^{\text{Field}}(z)$ is the results from \cite{Dominik2013} obtained from the Synthetic Universe website \footnote{The merger rates from \cite{Dominik2013} and several other studies using \texttt{StarTrack} are available from the Synthetic Universe website (\url{http://www.syntheticuniverse.org/})}.  

In the local universe, the standard model suggests that 15\% (12\%) of merging BBHs will have been formed in a GC under the high-end (low-end) metallicity scenarios.   For the optimistic model, the field rate increases significantly, with $F_{GC}(z)$ dropping to 4\% (2\%) in the optimistic case.  For the models using the delayed supernova prescription, the fraction increases to 25\% (15\%) in the local universe.

However, for the high-BH natal kick model, the merger rate from BBHs is comparable to those from the field, with $F_{GC}(z)$ reaching values of 55\% (45\%) in the local universe.  This surprising result is a direct implication of the dynamical formation scenario.  It was previously assumed that, were BHs to be formed with large natal kicks, the majority of BHs would be ejected from clusters, with only a small fraction (the low-velocity tail of the kick distribution) being retained and processed into BBHs.  Given that, it is surprising that rates from GCs should rival those from isolated binary stellar evolution in the field. 

To understand this, assume that there is some small fraction of BHs, $f_{\text{ret}}$, born with sufficiently low kicks that they remain gravitationally bound to whatever system they are member of, be it a GC or a binary star system.  For a GC, this means that after 20 Myr, approximately $N f_{\text{ret}}$ of the total $N$ BHs will be retained in the cluster.  If we then assume that some fraction, $f_{\text{BBH}}$ of these BHs will be dynamically processed into BBHs, then the total number of BBHs produced by the cluster is $N_{\text{BBH}} \propto N f_{\text{ret}}f_{\text{BBH}}$.  However, for systems in the field, each BBH must be formed from a binary progenitor, and therefore each binary must survive \emph{two} natal kicks.  From an initial population of $N_{\text{bin}}$ binaries, this suggests that the number of BBHs produced by the field in this scenario is $N_{\text{BBH}} \propto f_{\text{ret}}^2 N_{\text{bin}}$.  In other words, if the BH natal kicks are inversely proportional to the fraction of retained BHs, then as the kicks are increased, the rates from GCs decrease as $V_{\text{natal}}$, while the rates from the field decrease as $(V_{\text{natal}})^2$. 

Additionally, the retention of BHs in our high-kick GC model is aided by the primordial binary fraction.  Of the 27 BHs retained after supernova, 18 were in binaries at the time of formation.  As such, these systems are bound by both the local gravitational potential of their companions and the full potential of the cluster.  We note that our choice of initial binary fraction ($f_b = 10\%$) is much lower than the fraction assumed by \cite{Dominik2013}.  As a quick check, we rerun our high-kick GC model with a binary fraction of 50\%.  We find that the high-kick GC model now retains 48 BHs initially (41 of which were in binaries at formation), and creates 16 BBH mergers over its 12 Gyr lifetime.  Assuming these results scale with the total merger rate of BBHs from GCs, then roughly 75\% (70\%) of all BBH mergers in the local universe would have formed in GCs.  This approach is similar to proposed mechanisms for the retention of NSs in GCs \cite{Pfahl2002}.

Three caveats must be mentioned at this point.  First, it is unlikely that clusters, and particularly the massive GCs studied here, were formed with a binary fraction near 50\%.  Theoretical studies have shown that the fraction of binaries in a GC remains roughly constant over time \cite{Hurley2007,Ivanova2005}, while observations of GCs suggest a binary fraction between 1\% and 5\% \cite[e.g.~NGC 6397,][]{Davis2008}.  Second, increasing the binary fraction significantly changes the evolution of the GC as the primordial binaries heat the cluster core \cite[e.g.][]{Chatterjee2013}.  And finally, such a dependence on the binary fraction will also introduce a significant dependence on the initial conditions for binaries in star clusters, such as the distribution of initial mass ratios and semi-major axes.  We explore the effects of our choice of initial conditions in \cite{Chatterjee2016}.

\subsection{What is the field?}

In several papers examining the BBH merger rate of the field, it has been assumed that the field and star clusters exist as separate populations, formed by fundamentally different processes  However, observational evidence suggests that \emph{the majority of star formation occurs in clusters} \cite{Lada2003}.  These regions of star formation, consisting of clusters as small as $\sim 100M_{\odot}$, might dominate the star formation rate, particularly at high redshifts \cite{Kruijssen2015}.    Furthermore, observational evidence of the peculiar velocities of O stars in the MW suggest that $\sim 96\%$ of these massive stars formed in clusters \cite{DeWit2005}.  If that is the case, then a clean division between ``field'' and ``dynamical'' sources could be a significant oversimplification.  Studies of young star clusters ($\sim 3\times 10^3M_{\odot}$) have shown that the properties of a BBH population can be significantly altered by dynamics, even for small clusters that disrupt on a short ($\sim 100$ Myr) timescale \cite{Ziosi2014}.

A proper analysis of the contribution from clusters of all masses to the BBH
merger rate is beyond the scope of this paper; however, it is important to ask
the question:  what fraction of the merger rate of field binaries computed in population synthesis studies is actually affected by dynamics, which these studies neglect entirely?
As an order-of-magnitude check, let us assume that, consistent with our IMF,
approximately one of every 600 stars will become a BH.  If we assume that,
at the very least, a cluster must contain two BHs for dynamics to play a role,
then combined with an average stellar mass of $\left<m\right> = 0.6M_{\odot}$,
the minimum cluster mass where dynamics could be considered is $\sim 700
M{\odot}$.  If the fraction of star formation that occurs in clusters at high
redshifts is $\sim 50\%$ \cite[c.f.\ Figure 9,][]{Kruijssen2015}, and the cluster
initial mass function follows a $P(m) \propto 1/m^2$ distribution, then,
integrating from $100M_{\odot}$ to $10^8 M_{\odot}$ suggests that $\sim 10\%$ of
BBHs are formed in regions where dynamics may play a significant role.

Such an increase could play a significant role in the merger rate of dynamically-formed BBHs.  The results presented here have considered only the
present-day population of GCs which, at $z=0$, constitute only
$\sim 0.07\%$ of the baryonic matter in the MW \cite{PortegiesZwart2010}.  If most
galaxies formed with $\sim 10\%$ of stars in dynamically-relevant
clusters, then a significant fraction of the BBH population may have been affected by dynamics at some point.

However, this does \emph{not} directly imply that the BBH merger rate would be
dominated by clusters.  The majority of these disrupted clusters would have
masses significantly below the population of old, massive GCs studied here.  And
since we have shown that less-massive clusters not only produce fewer BBHs, but a
lower fraction of BBHs that will merge in a Hubble time, it is not immediately
obvious whether an early population of disrupted clusters will significantly affect the merger rates.  This is consistent with \cite{Ziosi2014}, which found that in young star clusters,
only 0.3\% of BBHs merged within 12 Gyr.

\section{Conclusion}

In this paper, we have explored mergers of BBHs from GCs.  Using our cluster Monte Carlo code, CMC, we have created a broad range of GC models with different masses, metallicities, and initial virial radii, designed to approximate the distribution of GCs observed in the present-day universe.  To increase the realism of our models, and to facilitate an easier comparison to BBH merger rates from the field (such as those from \cite{Dominik2013}), we have upgraded the stellar evolution prescription used by CMC, with new temperature-dependent stellar winds for O and B stars and new prescriptions for the supernova mechanism.  These changes were designed to bring our stellar evolution subroutines into agreement with those currently employed in the \texttt{StarTrack} population synthesis code.  With these enhancements, CMC now reproduces the mass distribution for single BHs that was used in the most recent estimates of BBH merger rates from the field \cite{Dominik2012,Dominik2013,Dominik2014}.

By considering a wide range of models, we were able to broadly characterize the relationship between the global properties of a GC (its mass and radius), and the inspiral times of the binaries it creates.  We showed that, in addition to creating more BBHs, more massive and more compact clusters ejected BBH binaries with higher binding energies and smaller semi-major axes, leading a greater number of BBHs to merge within 12 Gyr.  This explained the significant increase in merger rates first reported in \cite{Rodriguez2015a}: by accurately modeling the median and high-mass end of the GCMF, our realistic cluster models produced significantly more merging BBHs than previous studies of this type.  After integrating over the full GCMF, we found that the average GC will have produced 260 BBH mergers throughout its 12 Gyr lifetime.

We then used the GCMF to select a population of inspiral events representative of the population of GCs in the present-day universe.  With the new prescriptions for wind-driven mass loss and the rapid supernova mechanism, our GC models can now form BBHs with total masses from $20 M_{\odot}$ to $160M_{\odot}$.  These BBHs follow the same story that we first observed in \cite{Morscher2015,Rodriguez2015a}: the most-massive BHs lead the first period of mass segregation, driving the deep collapse of the cluster core.  These BHs then dynamically form BBHs, which are among the first to be ejected from the cluster, and the first to merge.  The GC processes through its BH population from most to least massive, continuing to eject BBHs up to the present day.  As such, the total mass of the merging BBHs decreases with redshift, with a median BBH total mass in the local universe of $\sim 40M_{\odot}$ and 50\% of sources lying between $\sim 35M_{\odot}$ and $\sim 50M_{\odot}$.  However, there exists a significant tail of massive sources, ejected with large semi-major axes (and correspondingly large inspiral times), such that the 90\% interval of masses extends from $\sim 30M_{\odot}$ up to $\sim 90M_{\odot}$.  We found that these massive BBHs were more likely to be formed in lower mass GCs, as these clusters ejected binaries with wider separations and longer inspiral times.  

We also found that the distribution of mass ratios depended strongly on the BH mass spectrum.  Any peaks in the distribution of single BH masses created corresponding peaks at twice that mass in our BBH population, where these BHs formed a large population of equal-mass BBHs.  As such, the three most prominent peaks in the BH mass distribution (and 18$M_{\odot}$, 30$M_{\odot}$, and 53$M_{\odot}$) created corresponding peaks in the distribution of BBH total masses at 36$M_{\odot}$, 60$M_{\odot}$, and 106$M_{\odot}$.  In between these regions, binaries would tend to form with unequal mass components, often drawing one of their components from the peaks of the BH mass spectrum.  This created an overabundance of sources with only one component from the peak of the BH mass spectrum, creating certain regions of the BBH total mass distribution (20, 45, and 80 $M_{\odot}$) that favored systems with unequal component masses.  

We then computed the merger rates from this population of BBHs as a function of
redshift.  In the local universe, we found that BBHs formed in GCs will merge at
a rate of $\sim 5$ Gpc$^{-3}$ yr$^{-1}$.  Under highly optimistic assumptions,
this rate becomes $\sim 20$ Gpc$^{-3}$ yr$^{-1}$, while highly pessimistic
assumptions forces the rate down to $\sim 2$ Gpc$^{-3}$ yr$^{-1}$.  For the
standard assumptions, we also found a merger rate of $\sim 3$ Gpc$^{-3}$
yr$^{-1}$ for sources with total masses $M < 40M_{\odot}$, and $\sim 2$
Gpc$^{-3}$ yr$^{-1}$ for sources with total masses $40M_{\odot} < M <
80M_{\odot}$.  This is consistent with the merger rate of $2-53$
Gpc$^{-3}\,\text{yr}^{-1}$ \cite{Abbott2016b} for binaries similar to GW150914.  When comparing these numbers to estimates of BBH merger rates from the field, we found that roughly 15\% of BBH mergers in the local universe will have originated in a GC.  However, if BHs are born with large natal kicks, similar to NSs, this fraction increased significantly, with GCs possibly dominating the merger rate for BBHs.  In addition to changing BBH merger rates and properties, any such changes effecting BH retention in early GCs (such as the natal kicks or the high-mass slope of the IMF) will significantly change the observational properties of GCs in the present day.  A study to fully characterize the relationship between BH retention and GC observational properties is currently underway \cite{Chatterjee2016}.

We reiterate that we have studied a conceptually clean, but not physically complete, picture of star formation and BBH formation in the universe.  By using the present-day population of observed GCs, we have restricted ourselves to studying a well-posed problem with good observational constraints.  However, both the current study and studies of field populations have ignored the fact that \emph{most} star formation is assumed to occur in clusters, with lower-mass clusters being disrupted into what we today call ``the field''.  The merger rates and BBH properties of estimates from the field should only be applied to binary systems from star forming regions that were disrupted before they could be dynamically altered; however, it is not at all obvious that the results presented here can simply be scaled down to lower-mass star forming regions.  A full exploration of the BBH merger problem will require an exploration of the intermediate regime, where most BBHs are formed from primordial binaries that undergo significant dynamical perturbations (e.g. \cite{Ziosi2014}).  Fully understanding these three scenarios--the field, dynamical, and intermediate BBH formation regimes--will be critical to unlocking the potential of gravitational-wave astrophysics.

Finally, we have shown that the masses and redshift associated with the recent detection of GW150914 (and the less-significant GW trigger LVT151012) are consistent with dynamical formation in the low-metallicity environment of a GC as reported in our models.  However, we will address the full implications of this detection in the context of GC modeling in a future work \cite{Rodriguez2016c}.  This detection emphasizes the importance of BBH modeling as we transition into the era of GW astrophysics.

\begin{acknowledgments}
We would like to thank Jarrod Hurley, Chris Belczynski, Vicky Kalogera, Onno Pols, Francesca Valsecchi, and Cole Miller for useful discussions.
CR was supported by an NSF GRFP Fellowship, award DGE-0824162.  This work was supported by NSF Grant AST-1312945 and NASA Grant NNX14AP92G.
\end{acknowledgments}

\bibliography{mendleBlob}

\end{document}